\documentclass[aps,prb,twocolumn,superscriptaddress,showpacs,a4paper,10pt]{revtex4-2}
\usepackage{graphicx}
\usepackage{dcolumn}
\usepackage{bm}
\usepackage{ulem}
\usepackage{hyperref}
\usepackage{comment}
\usepackage[dvipsnames]{xcolor}
\usepackage{chemformula}
\usepackage{braket}

\usepackage{multirow}

\begin{document}
	\title{
How orbitals and oxidation states determine apparent topographies in
scanning tunneling microscopy: the case of fluorine on silver
surfaces
 }
 

\author{Adri\'{a}n G\'{o}mez Pueyo}
\affiliation{ISC-CNR, Istituto dei Sistemi Complessi, via dei Taurini 19, 00185 Rome, Italy}%
\affiliation{Dipartimento di Fisica, Sapienza Università di Roma, 00185 Rome, Italy
}

\author{Jazm\'{i}n Arag\'{o}n S\'{a}nchez}
 	\affiliation{Dipartimento di Fisica, Universit\`a di Roma ``Tor Vergata'', via della Ricerca Scientifica, 1-00133 Rome, Italy}

\author{Ilya Degtev}
\affiliation{ISC-CNR, Istituto dei Sistemi Complessi, via dei Taurini 19, 00185 Rome, Italy}%
\affiliation{Dipartimento di Fisica, Sapienza Università di Roma, 00185 Rome, Italy
}

\author{Maria Eleonora Temperini}%
\affiliation{Dipartimento di Fisica, Sapienza Università di Roma, 00185 Rome, Italy
}

\author{Daniel Jezierski}
 \affiliation{University of Warsaw, Center of New Technologies, Żwirki i Wigury 93, 02-089 Warsaw, Poland}

\author{Conor Hogan}
\affiliation{ISM-CNR, Istituto di Struttura della Materia, 00133 Roma, Italy}
\affiliation{Dipartimento di Fisica, Universit\`a di Roma ``Tor Vergata'', via della Ricerca Scientifica, 1-00133 Rome, Italy}

\author{Antonio Caporale}%
 \affiliation{Dipartimento di Fisica, Universit\`a di Roma ``Tor Vergata'', via della Ricerca Scientifica, 1-00133 Rome, Italy}

\author{Luciana Di Gaspare}
\affiliation{Dipartimento di Scienze, Università Roma Tre, 
Viale Guglielmo Marconi 446, 00146 Rome, Italy}

\author{Luca Persichetti}
\affiliation{Dipartimento di Fisica, Universit\`a di Roma ``Tor Vergata'', via della Ricerca Scientifica, 1-00133 Rome, Italy}

\author{Monica De Seta}
\affiliation{Dipartimento di Scienze, Università Roma Tre, 
Viale Guglielmo Marconi 446, 00146 Rome, Italy}

\author{Wojciech Grochala}%
 \affiliation{University of Warsaw, Center of New Technologies, Żwirki i Wigury 93, 02-089 Warsaw, Poland}

\author{Paolo Barone}
\affiliation{SPIN-CNR, Istituto Superconduttori, Materiali Innovativi e Dispositivi, Area della Ricerca di Tor Vergata, via del Fosso del Cavaliere 100, 00133 Rome, Italy}	

\author{Luca Camilli}
 \email{luca.camilli@roma2.infn.it}
\affiliation{Dipartimento di Fisica, Universit\`a di Roma ``Tor Vergata'', via della Ricerca Scientifica, 1-00133 Rome, Italy}

  \author{Jos\'{e} Lorenzana}
 \email{jose.lorenzana@cnr.it}
\affiliation{ISC-CNR, Istituto dei Sistemi Complessi, via dei Taurini 19, 00185 Rome, Italy}%
\affiliation{Dipartimento di Fisica, Sapienza Università di Roma, 00185 Rome, Italy
}


\date{\today}
	
\begin{abstract}
We use density functional theory calculations to characterize the early stages of fluorination of silver's (100) and (110) surfaces. In the Ag(100) surface, the hollow site is the most favorable for F adatoms. In the Ag(110) surface, three adsorption sites, namely hollow, long bridge, and short bridge, exhibit similar energies. These locations are also more favorable than an F adatom occupying a vacancy site irrespectively of whether the vacancy was present or not in the pristine surface. The computed energy as a function of surface coverage is used to compute the equilibrium thermodynamics phase diagram. We argue that for the typical pressure and temperature of fluorination experiments, the state of the surface is not determined by thermodynamics but by kinetics. 
Combining these results with scanning tunneling microscopy (STM) topographic simulations we propose assignments to features observed experimentally. We present a minimal model of the apparent topography of adatoms in different locations in terms of hydrogenic orbitals, explaining the observed trends. The model links the STM apparent topography to structural information and the oxidation states of the Ag atoms near the adatom.  
\end{abstract}
	
\maketitle
	
	
    \section{\label{sec:Introduction}Introduction}
The interaction of halogens, chalcogens, and other adatoms with metallic surfaces is of fundamental importance in several fields such as electrochemistry~\cite{Roman2014}, corrosion,  heterogeneous catalysis~\cite{Besenbacher1993,Andryushechkin2018b, Lin2021} and opens up the possibility to synthesize exotic quantum materials such as, for example, spin-1/2 silver fluorides ~\cite{Grochala2001,McLain2006,Grochala2006a,Yang2014,Gawraczynski2019,Grzelak2020,Miller2020,Sanchez-Movellan2021,Bachar2022,Piombo2022,Prosnikov2022,Wilkinson2023}.

Density functional theory (DFT)~\cite{Ignaczak1997,Tripkovic2009,Roman2014,Zhu2016,Zaum2018,Spurgeon2019,Lee2020} is an invaluable theoretical tool to study atom adsorption on metal surfaces. DFT computations provide insights into key parameters, including the distance between the adatom and the upper metal layer, the proximity to the nearest metal atom, charge distributions among the atoms, changes in the work function induced by the adsorbate, and the energy associated with adsorption which is essential for thermodynamic considerations~\cite{Li2002a}.

From the experimental point of view, scanning tunneling microscopy (STM) provides unique information on how adatoms interact with metallic surfaces. STM provides an ``apparent topography" (AT) of the surface.  This, however, may be different from naive expectations based on the charge distribution.
For example, operating in constant current mode, 
sometimes adatoms appear as protrusions but a depression is also possible, as well as a combination of both yielding a Mexican hat or ``sombrero" shape. For example, sombrero shapes are observed in the case of sulfur~\cite{Spurgeon2019}, and oxygen~\cite{Schintke2001} on silver. More recently, sombrero shapes have been observed for fluorine on silver surfaces~\cite{Sanchez2024}. This variety of ATs is due to the fact that the conductance depends on several factors, such as the density of states and the extension of orbitals. Therefore, modeling the STM data is mandatory to extract all the physical and chemical information from the STM results.  

Early theories~\cite{Lang1986,Lang1987} of the ATs of adsorbed atoms on metallic surfaces were based on the jellium model. This provided insights into the bias dependence of the tunneling current and the apparent height of the adatom right at its center but did not give information on the apparent topography away from the center, i.e., what kind of feature the adatom produces. A more complete theory~\cite{Sautet1997} in terms of a tight-binding model provided more detailed information on the ATs and predicted protrusions, depressions, and the sombrero shape depending on the adatom. DFT can compute~\cite{Spurgeon2019} STM ATs based on the Tersoff-Hamann approximation~\cite{Tersoff1983,OTERODELAROZA2009,OTERODELAROZA20141007} thus allowing the identification of the features observed in the experiment.

In Ref.~\cite{Sanchez2024} low-temperature (LT)-STM measurements have been performed on the Ag(100) and Ag(110) surfaces of single crystals exposed to fluorine and hydrogen fluoride (HF). Here we present DFT computations to identify the energetically most favorable adatom locations and the corresponding ATs. We consider both adatoms on top of the ideal surfaces and adatoms in surface vacancy sites. We use the DFT results to estimate the range of fluorine gas chemical potential at which a given coverage is thermodynamically stable and compare it with the condition to have stable bulk phases. 

The energetics at small coverage allow us to propose assignments of the experimentally observed ATs to specific high-symmetry locations on the surfaces. This is complemented with computations of the ATs of the most stable adatom locations. Experimental ATs have a rich internal structure. As mentioned, the  Ag(100) and some Ag(110) ATs show a sombrero shape, especially for negative voltage biases. Instead, ATs acquired with positive voltages tend to show a depression shape of the adatom. Simulated ATs in DFT generally reproduce this behavior although some discrepancies remain. 

As well known, DFT provides a wealth of information by including all relevant orbitals and a realistic effective potential. However, the trade-off is that it is not always straightforward to pinpoint the important factors determining the AT in terms of the chemistry and structure of the surface. To obtain a deeper understanding of the AT, we present a simplified orbital model of the STM topographies. This allows us to link the main observed features in STM data to information such as the out-of-plane distance from the Ag surface of the F adatom and the oxidation state of the nearby Ag atoms.
	
\begin{figure}[tttt]
    \begin{center}			\includegraphics[width=\columnwidth]{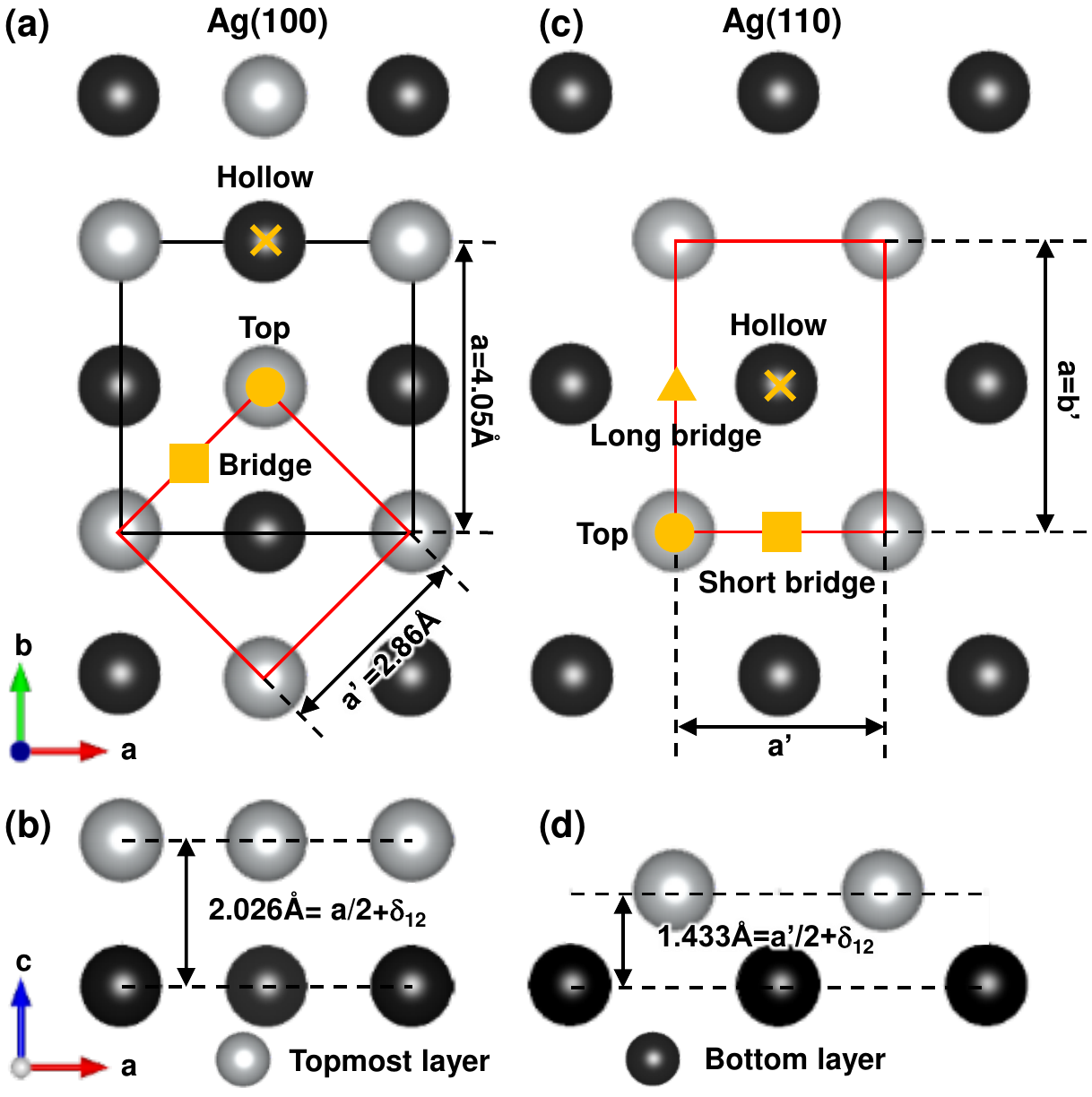}
    \end{center}
    \caption{Geometry of the first two layers of the (a)-(b) Ag(100) and (c)-(d) Ag(110) slabs.
    The square and rectangle drawn with red solid lines in (a) and (c) show the unit cell considered for each surface. The crystallographic axis on the left indicates the orientation of the bulk crystal associated to the Ag(100) surface. For the 110 surface shown in (c) the horizontal direction corresponds to [110] while the vertical direction corresponds to [001]. 
    The highly-symmetric positions are indicated with yellow shapes for both surfaces.  
    The light and dark gray spheres represent silver atoms from the topmost and the underlying layers, respectively. The distance between these two layers is indicated with an arrow in panels (b) and (d). The parameter $\delta_{12}$ characterizes the interlayer distance relaxation for the two topmost layers.} \label{Fig:surf_param}
\end{figure}

This work is organized as follows. The methods used are explained in Sec.~\ref{sec:Methods}. Structural parameters and the energetics and diffusion barriers of adatoms are discussed in Sec.~\ref{sec:structure}. The thermodynamics of a silver surface exposed to fluorine gas is discussed in Sec.~\ref{sec:thermoresults}. Results for the topographies of adatoms are discussed in Sec.~\ref{sec:topographies}. Section~\ref{sec:modelresults} deals with the explanation of the topographic features as the sombrero shape in terms of a simplified orbital model. The summary and conclusions of the work can be found in Sec.~\ref{sec:Conclusions}.

    \section{\label{sec:Methods}Methods}
\subsection{Surface structure and definitions}
  
For Ag(100) the atoms in the surface present a square unit cell with lateral dimension $a'=a/\sqrt2$ and area $(a')^2$, with $a$ the bulk lattice constant. There are three highly symmetric sites for the adsorption of the fluorine atoms [Fig.~\ref{Fig:surf_param}\,(a)]: hollow, bridge, and top site. The dimensions shown in the figure were obtained from the DFT computations as explained below. 

The Ag(110) surface has a similar layout albeit with a rectangular unit cell of area $a'\times b'$, with $b'=a$. Consequently, two types of bridge sites are present: a long-bridge site, positioned between two Ag atoms along one of the longer sides of the rectangular lattice, and a short-bridge site, located between two Ag atoms along one of the shorter sides. This configuration yields a total of four highly symmetric sites suitable for the adsorption of fluorine atoms in ideal surfaces, as shown in Fig.~\ref{Fig:surf_param}\,(c). 
   
We define the surface coverage as $\Theta=N_{\ch{F}}/A$ (ML) where $N_{\ch{F}}$ is the number of fluorine adatoms on the surface and $A$ is the surface area in units of the surface unit cell. For both surfaces, $A$ corresponds to the number of each type of highly-symmetric sites in the pristine surface, and $\Theta=1$ ML corresponds to the full occupation of all the available sites of one kind [considering bridges along the two perpendicular directions in the Ag(100) surface as distinct].

\subsection{Characterization of Adsorption}

The surface can be exposed to monoatomic fluorine~\cite{Sanchez2024} or diatomic fluorine depending on experimental conditions.

If fluorination proceeds from exposing the clean Ag surface to the diatomic gas, the molecule must be dissociated during the reaction process. The zero-temperature dissociation energy per atom reads,
\begin{equation}
  \label{eq:ded}
  E_d\equiv E({\ch{F}})-E({\ch{F2}})/2,
\end{equation}
where $E({\ch{F2}})$ being the energy of a \ch{F2} molecule and $E({\ch{F}})$ the energy of a
 single F atom.
 At zero temperature, these correspond to the chemical potentials 
of the molecular ($m$)  and atomic ($a$) gas namely,  
 \begin{eqnarray}
  \mu_{\ch{F}}^{0(m)}&\equiv& E({\ch{F2}})/2, \ \ \ \ \ (\ch{F2})\label{eq:mu0m}\\
  \mu_{\ch{F}}^{0(a)}&\equiv& E({\ch{F}}). \label{eq:mu0a}\ \ \ \ \ (\ch{F})
\end{eqnarray}

We define the adsorption energy per atom as,
\begin{equation}
  E_{ad}^{(g)}(\Theta)= (E_{\ch{Ag:F}}-E_{\ch{Ag}}-\Delta N_{\ch{Ag}}\mu^0_{\ch{Ag}}
  -N_{\ch{F}}\mu_{\ch{F}}^{0(g)})/N_{\ch{F}}, \label{eq:em2}
\end{equation}
where $g=a,m$ depending on the kind of gas exposure. 
$E_{\ch{Ag:F}}$ and  $E_{\ch{Ag}}$ 
are the total energy of the fluorinated system and the pure silver slab respectively. $\mu^0_{\ch{Ag}}$ corresponds to the  energy per atom of bulk Ag. These are the total energies of fully relaxed systems. $N_{\ch{F}}$ is the number of \ch{F} atoms in the slab. $\Delta N_{\ch{Ag}}$ takes into account that the number of Ag atoms may change during the adsorption process. For example, a silver atom may be kicked off the surface and replaced by an F atom (exchange reaction). Thus, $\Delta N_{\ch{Ag}}=(N_{\ch{Ag}}-N^0_{\ch{Ag}})$ is the difference of Ag atoms in the fluorinated slab $(N_{\ch{Ag}})$ and the pristine slab ($N_{\ch{Ag}}^0$). With these definitions $E_{ad}^{(g)}<0$ indicates an exothermic reaction. 

Often experiments and theoretical works assume that Ag is exposed to \ch{F2} gas, and therefore $E_{ad}^{(m)}$ is usually computed.
The two adsorption energies are related by, 
\begin{equation}
    \label{eq:eda2edm}
    E_{ad}^{(a)}=E_{ad}^{(m)}-E_d. 
\end{equation}

\subsection{Thermodynamics}\label{sec:thermo}

We can use the DFT results to estimate the equilibrium pressures to stabilize the various coverages. 
We follow Ref.~\onlinecite{Li2003a} and consider an Ag surface in contact with \ch{F2} gas or \ch{F} gas. We want to compute the equilibrium coverage $\Theta$ as a function of temperature $T$ and pressure $p$. For this, we consider a constrained Gibbs free energy of the whole system (bulk+surface+gas), $G_{Ag:F}(T,p,N_{Ag},N_F,\Theta)$, where we use $\Theta$ as an additional thermodynamic variable. The equilibrium $\Theta$ for each pressure of the gas can be obtained by minimizing $G_{Ag:F}$. 


For a system with only one kind of surface, we can characterize the surface Gibbs free energy with a single parameter,
\begin{eqnarray}
    \gamma(T,p,\Theta)&&=\\
&&    \frac{G_{Ag:F}(T,p,\Theta)-N_{Ag} \mu_{Ag}(T,p)-N_F\mu_F(T,p)}A,\nonumber
\end{eqnarray}
where $\mu_{Ag}(T,p)$ and $\mu_F(T,p)$ are the chemical potentials of the silver bulk and the gas. 

It is useful to write the gas chemical potential as a $T=0$ contribution plus a correction, 
\begin{equation}
\label{eq:muf}
  \mu_F(T,p)=\mu_F^{0} +\Delta \mu_F(T,p).
\end{equation}

We will assume that the surface phases are ordered and that the contribution of the solid to the Gibbs free energy can be computed~\cite{Li2002a,Li2003a,Reuter2001} at $p=T=0$ with DFT. Then all the pressure and temperature dependence will come from $\Delta \mu_F(T,p)$. Therefore, for a slab geometry with an unrelaxed surface, we write,
\begin{equation}
    \gamma(T,p,\Theta)=[E_{Ag:F}-N_{Ag} \mu_{Ag}^0-N_F\mu_F(T,p)]/A-\sigma^u.
\end{equation}
Here we took into account that in the slab geometry there are two non-equivalent surfaces introducing the surface energy of the unrelaxed surface $\sigma^u$. 

Subtracting the $\Theta=0$ contribution we obtain,
\begin{equation}
    \label{eq:dgdtp}
\Delta\gamma^{(g)}(T,p,\Theta)=\Delta\gamma^{(g)}(0,\Theta)-\Theta \Delta \mu_F^{(g)}(T,p),
\end{equation}
where we added the $g$ label to distinguish the kind of gas exposure and 
\begin{equation}\label{eq:dgdtp0}
    \Delta\gamma^{(g)}(0,\Theta)=  \Theta E_{ad}^{(g)}(\Theta). 
\end{equation}
    
The pressure dependence of the chemical potential can be approximated by the ideal gas contribution ($k_BT \ln(p/p_r)$) with respect to a reference pressure $p_r$. Therefore $\Delta\mu_F^{(a)} (T,p)$ separates into a temperature dependence $\widetilde{\mu}(T,p_r)$ and a pressure dependence, namely, 
\begin{eqnarray}
  \label{eq:muf2}
\Delta\mu_F^{(a)} (T,p)&=&\widetilde{\mu}_F(T,p_r)+k_BT \ln(p/p_r),  \ \ \ \ \ (\ch{F})\\
  \Delta\mu_{F}^{(m)}(T,p)&=& \frac{\widetilde{\mu}_{F_2}(T,p_r)+k_BT \ln(p/p_r)}2.\ \ \  (\ch{F2})\
\end{eqnarray}
For monoatomic fluorine the temperature dependence can be estimated from the ideal gas expression, 
\begin{equation}\label{eq:ig}
    \widetilde{\mu}_F(T,p_r)=k_B T \ln \left[\frac{\hbar^3 p_r }{(k_b T)^{5/2}}\left(\frac{2\pi}M\right)^{3/2}\right].
\end{equation}
For $\widetilde{\mu}_{F_2}(T,p_r)$ one should take into account rotational and vibrational contributions. Alternatively one can use tabulated experimental data for both.

\subsection{\label{subsec:Computational}Computational details}

The structural and electronic properties of F atoms adsorbed on Ag(100) and Ag(110) surfaces were studied using DFT as implemented in the Vienna Ab initio Simulation Package~\cite{VASP1,VASP2,VASP3,VASPPAW} (VASP). All the calculations were done using the PBEsol~\cite{Perdew2008} exchange and correlation functional. The energy cutoff for the plane-wave basis used was 700 eV. Gaussian smearing was chosen to determine the partial occupations of each orbital, with a value for the width of the smearing of 0.05 eV. The electronic self-consistent loop was considered converged when the energy difference between two steps was smaller than $10^{-6}$ eV.

As a starting point, we performed a relaxation of the bulk Ag unit cell using a $32\times 32\times 32$ $k-$point $\Gamma$-centered mesh. We used the conjugate-gradient algorithm as implemented in VASP to minimize the forces among ions until the energy difference between two ionic steps was smaller than $10^{-5}$ eV. We found a lattice constant of $a=4.05$ \AA{} for bulk Ag, which agrees well with the experimental value of 4.08 \AA{}~\cite{King1981,Hu2005}. 
    
In order to study surface effects we constructed slabs of Ag using the bulk Ag cell. This leads to a surface unit cell $a'=a/\sqrt2=2.86$ \AA{} for the Ag(100) surface and  $a'=a/\sqrt2= 2.86$ \AA, $b'=a=4.05$ \AA{} for the Ag(110) surface. 

The slabs were composed of 8 layers of Ag and a vacuum space corresponding to 13 times the interlayer separation of each surface, $\approx26.3$ \AA\ for the Ag(100) surface and $\approx18.6$~\AA\  for the Ag(110). The four top layers of Ag were free to relax while the four bottom layers remained fixed, simulating the bulk of the system. The interlayer distance, $\delta_{ij}$, between two consecutive layers $i$ and $j$ was computed from the relaxed structures. Figures~\ref{Fig:surf_param} (b) and ~\ref{Fig:surf_param} (d) illustrate the definition of the relaxation $\delta_{12}$ for the two topmost layers and Table~\ref{Table:layer_dist} reports the parameters for all relaxed layers. These values compare well with those found in previous works for the Ag(100)~\cite{Perdew20172} and Ag(110) surfaces~\cite{Perdew20172,Banerjee2018}.

\begin{table}[t]
    \caption{Relative height change between contiguous pairs of layers of the relaxed part of the Ag(100) and Ag(110) slabs with respect to the bulk.
    $\delta_{ij}$ denotes the difference between the distance of the $i$ and $j$ layers and the interlayer separation of the bulk.
    The values in parenthesis correspond to the percentage change.} \label{Table:layer_dist}
\begin{ruledtabular}
    \begin{tabular}{ccc}
{}      & {Ag(100)} & {Ag(110)} \\
      \hline
      {$\delta_{12}$ (pm)} & -3.6  (-1.8\%) & -12.3  (-8.6\%)\\
      {$\delta_{23}$ (pm)} & 0.7  (0.3\%) & 6.1 (4.3\%)\\
      {$\delta_{34}$ (pm)} & 0.6  (0.3\%) & -2.1  (-1.5\%)\\
    \end{tabular}
\end{ruledtabular}
\end{table}

To study different concentrations of F on the surface we constructed supercells of the slabs. For the fluorinated Ag(100) surface, $2\times2$ and $2\sqrt{2}\times 2\sqrt{2}$ (in units of $a'\times a'$) supercells were prepared and a $16\times 16\times 1$ $k-$point $\Gamma$-centered grid was used, while for the Ag(110) we used $1\times2$ and $4\times 2$ supercells (in units of $a'\times b'$) with $33\times 11\times 1$ and $8\times 11\times 1$ meshes, respectively. The built-in dipole correction~\cite{Neugebauer1992,Makov1995} along the direction perpendicular to the surface was applied to avoid errors derived from applying periodic boundary conditions to a slab with a net dipole moment.

We then added the F adatoms on the highly symmetric positions discussed earlier, changing the number of adsorbates to sample different coverage values. Each calculation was carried out with a certain coverage of F atoms in the same type of site. Thus, we do not consider the possible interaction between adatoms in non-equivalent positions
due to the high computational cost. 

The four top layers of the slab were relaxed again with Ag atoms allowed to move in all directions and F atoms constrained along the axis perpendicular to the surface to ensure they remained in the highly symmetric positions we were interested in. Once we obtained the relaxed structure, the ground state (GS) energy was computed again. 
    
For the simulation of the STM topographies, we created large supercells that could accommodate the size of the features observed in the experiments, which was approximately 1 nm, and relaxed the four top layers and one F atom, like in the previous calculations. Once we had the relaxed slab with the adatom in the position that we wanted to study, we ran a GS calculation with a sparse $k$-point grid. Using the results from this calculation as a starting point, we calculated the density of states (DOS) and the STM topographies with a denser $k$-point grid, and then we processed these results using the Tersoff-Hamann approximation~\cite{Tersoff1983} as implemented in the open-source code Critic2~\cite{OTERODELAROZA2009,OTERODELAROZA20141007}. For the Ag(100) surface we used a $3\sqrt2\times3\sqrt2$ supercell with a $12\times 12\times 1$ $k-$point $\Gamma$-centered mesh~\cite{Migani2006} for the relaxation and GS calculations. The DOS and STM calculations were carried out using a $24\times 24\times 1$ $k-$point mesh. For the Ag(110) surface we used a $5\times4$ supercell with an $8\times 8\times 1$ $k-$point $\Gamma$-centered mesh~\cite{Migani2006} for the relaxation and GS calculations, while the DOS and STM calculations were carried out using a $16\times 16\times 1$ $k-$point mesh. In all STM computations, we included non-spherical contributions related to the gradient of the density in the projector-augmented-waves~\cite{Blochl1994} spheres to get accurate wave functions. We sampled values of the bias voltage, $V_{\rm B}$, of the tip of the microscope ranging from 3 to -3 V with steps of 0.25 V for both surfaces and tunneling currents, $I_{\rm T}$, ranging from 6.62 pA to 66 $\mu$A ($10^{-9}$ to $10^{-2}$ a.u. of current).

\subsection{A simplified orbital model of the STM topography\label{sec:orbmodel}}

To obtain an intuitive understanding of the origin of the different topographic features we will use a simplified orbital model presented here. We start by decomposing the tunneling current into distinct ``channels". Formally, for an $s$-wave tip in the Tersoff–Hamann approximation~\cite{Tersoff1983}, the tunneling current for a given bias voltage is given by,
\begin{equation}\label{eq:bardeen}
    I(\bm r,V_{\rm B})=\frac{4\pi e}\hbar|M|^{2}N_t(0)\int_0^{V_{\rm B}}  N(\bm r ,\epsilon) d\epsilon,
\end{equation}
where $N(\bm r,\epsilon)$ is the sample local DOS and $\bm r$ is the position of the center of the radius of curvature of the tip. $N_t(0)$ is the tip DOS and $M$ is a tip-dependent matrix element, both of them assumed to be energy independent. The DOS can be written in terms of the single particle eigenvalues, $\epsilon_{n}$, and eigenvectors, $\psi_{n}(\bm r)$, as, 
\begin{equation}
    N(\bm r ,\epsilon)=\sum_{ n }\psi_{n}^*(\bm r)\psi_{n}(\bm r)\delta(\epsilon-\epsilon_{n}).
\end{equation}

We assume the relevant single particle states can be described with a lattice model with atom positions $\bm R$ and orbitals  $\ket{\bm R lm}$, where $l$ labels point group irreducible representations and $m$ labels orbitals with the same $l$. If we neglect crystal field effects $l$ labels the angular momentum and principal quantum number of orbitals. We thus decompose the single-particle states, $\psi_n(\bm r)=\braket{\bm r|n}$, in the basis of Wannier functions, $\braket{\bm r|\bm R lm}=w_{\bm R lm}(\bm r)$, as, 
\begin{equation}
\psi_{n}(\bm r)=\sum_{\bm R lm} \braket{\bm R lm|n} w_{\bm R lm}(\bm r),\nonumber
\end{equation}
with coefficients $\braket{\bm R lm|n}$ obtained solving the lattice model. Similar to Refs.~\cite{DellAnna2005,Kreisel2016}, we write the local DOS as
\begin{eqnarray*}
    N(\bm r ,\epsilon)= \sum_{\bm R\bm R' ll'mm'n} \braket{n|\bm R lm}  &&   \braket{\bm R' l'm'|n}\delta(\epsilon-\epsilon_{n}) \\
 &&
\times   w_{\bm R lm}^*(\bm r)w_{\bm R' l'm'}(\bm r).
\end{eqnarray*}

The largest contributions arise from diagonal terms proportional to the Wannier orbital density. Thus, in a first approximation, we neglect all off-diagonal contributions, 
\begin{eqnarray}\label{eq:approxdiag}
    &&N(\bm r ,\epsilon)\approx \nonumber\\
    &&\sum_{\bm R lm n} \braket{n|\bm R lm}    \braket{\bm R lm|n}\delta(\epsilon-\epsilon_{n}) w_{\bm R lm}^*(\bm r)w_{\bm R lm}(\bm r).
\end{eqnarray}
For symmetry-related orbitals the orbital projected DOS (PDOS), defined as
\begin{equation}\label{eq:pdos}
    D_{\bm R l}(\epsilon)=  \sum_{n }  \braket{n|\bm R lm}  \braket{\bm R lm|n}  \delta(\epsilon-\epsilon_{n}),
\end{equation}
is independent of $m$. Thus the local DOS at the tip can be written as, 
\begin{equation}\label{eq:Rxdos}
    N(\bm r ,\epsilon)\approx\sum_{\bm R l}  D_{\bm R l} (\epsilon) \rho_{\bm R l}(\bm r),
\end{equation}
where 
\begin{equation}
\rho_{\bm R l}(\bm r)=\sum_{m}  w_{\bm R lm}^*(\bm r)w_{\bm R lm}(\bm r),\nonumber
\end{equation}
is the total charge density associated with the ``shell" $l$. Equation~\eqref{eq:Rxdos} allows the decomposition of the sample DOS entering into the tunneling current into contributions from different orbitals. These expressions will be used below to obtain a simple understanding of the apparent STM topography and gain insight into how it depends on structural and electronic characteristics.
 
    \section{\label{sec:TheoreticalResults}Results and Discussion}
\subsection{\label{sec:structure}Structure, stability and diffusion of adatoms}


Table~\ref{Table:atom_dist} reports the calculated height of the F adatoms and their nearest Ag neighbor(s) relative to the surface of the clean silver slabs for all the highly symmetric sites of the Ag(100) and Ag(110) surfaces, as well as for the case where one F adatom occupies a vacancy on the clean silver surface. Notice that the bridge of the Ag(100) surface and the short bridge of the Ag(110) surface correspond to the same Ag-Ag distance $a'$. Indeed, the F height $\approx$160 pm is very similar in both cases. We also report the charge transfer $\delta n$ between F and the nearest neighbor Ag atom on the surface.


We computed the adsorption energy, $E_{ad}$, for different levels of fluorine coverage for each type of highly symmetric site on both Ag(100) and Ag(110) surfaces. We report both the molecular and atomic adsorption energies, which differ by the theoretical value of $E_d=1.49$~eV [Eq.\eqref{eq:eda2edm}].

\begin{table}[tb]
  \caption{Electron charge transfer, $\delta n$, and height $d$ of the fluorine adatom and its silver nearest neighbor on the surface. $d$ is measured relative to the (relaxed) topmost layer of the clean Ag(100) and Ag(110) slabs for each of the highly symmetric sites studied. $\delta n$ for fluorine is defined as the difference in electron number relative to an isolated atom. For silver, the difference is relative to an atom of the pristine surface.
  Electron number are taken in spheres of radius 0.704~\AA, 1.503~\AA\, for F and Ag respectively. 
  S/L stands for short/long. 
  Calculations correspond to one F atom on the $3\sqrt{2}\times3\sqrt{2}$ and $5\times4$ supercells of the Ag(100) and Ag(110) surfaces respectively. 
  } \label{Table:atom_dist}
\begin{ruledtabular}
\centering
\label{tab:paired_columns}
\begin{tabular}{c|cc|cc|cc|cc}
\multirow{2}{*}{ } & \multicolumn{4}{c|}{Ag(100)} & \multicolumn{4}{c}{Ag(110)} \\
\multirow{2}{*}{} & \multicolumn{2}{c}{F} & \multicolumn{2}{c|}{Ag} & \multicolumn{2}{c}{F} & \multicolumn{2}{c}{Ag} \\
& $d$  & $\delta n$  & $d$  & $\delta n$  & $d$  & $\delta n$  & $d$  & $\delta n$ \\
Site &  (pm) &   &  (pm) &  &  (pm) &  & (pm) & \\
\hline
Hollow       & 123.7 & 0.185 & 1.6  & -0.019 & 94.9  & 0.173 &  2.0 & -0.028\\
Top          & 205.6 & 0.185 & 1.7  & 0.025  & 200.5 & 0.163 & -1.0 & 0.021\\
Bridge       & 160.7 & 0.197 & 6.0  & -0.016 & -     &  -    & -    & -\\
S bridge     & -     & -     & -    & -      & 163.8 & 0.199 & 5.4  & -0.020\\
L bridge     & -     & -     & -    & -      & 112.4 & 0.188 & 3.7  & -0.026\\
Vacancy      & -15.7 & 0.155 & -2.9 & -0.028 & -28.8 & 0.163 & -2.0 & -0.019\\
\end{tabular}
\end{ruledtabular}
\end{table}

It is well known that PBEsol overbinds diatomic molecules~\cite{Perdew1996}. Indeed, the experimental dissociation energy for fluorine~\cite{Matthiasson2021,Decorpo1970}, $E^{\rm exp}_d=0.815$ eV is much lower \footnote{Since our computation does not take into account the zero-point vibrational energy, it is more fair to remove the vibrational contribution~\cite{Pople1989} from the experiment, which yields $E^{\rm exp}_d=0.855$ eV, still significantly smaller than the theory} than the theoretical value  $E_d=1.49$~eV. This issue is particularly problematic for oxygen adsorption on Ag(111)~\cite{Li2002a} where only for very small coverage the theoretical adsorption energy is negative (exothermic) in relation to \ch{O2}. However, for \ch{F2} the problem is less severe, and we find that the adsorption energy remains negative even at full coverage (c.f. Fig.~\ref{Fig:Eads}, right scale). In the case of exposure to the monoatomic gas, the adsorption energies do not involve the dissociation energy, and this problem does not arise.

\subsubsection{Ag(100) surface}

\begin{figure}[tb]
    \begin{center}    \includegraphics[width=\columnwidth]{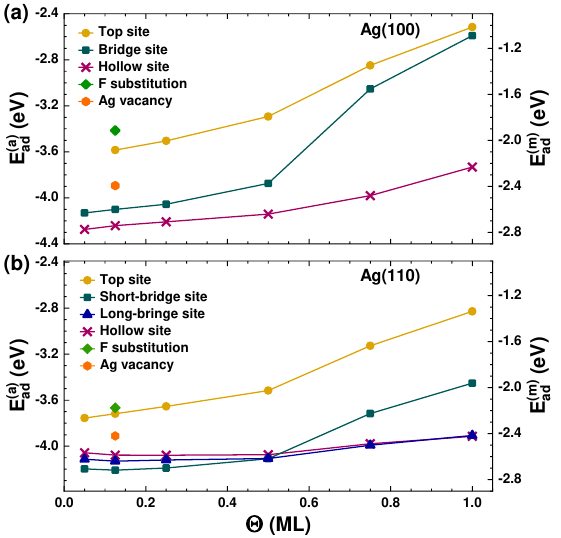}
    \end{center}
    \caption{(a)
    Adsorption energies per atom for atomic fluorine (left scale) and molecular fluorine (right scale),  as a function of the surface coverage, $\Theta$, of the Ag(100) surface for the three possible highly-symmetric positions: hollow, bridge, and top.
    (b) The same plot for the Ag(110) surface, where we have four highly-symmetric positions: hollow, long bridge, short bridge, and top. The values labeled as ``F substitution" correspond to an exchange reaction where an Ag atom is pulled from the surface and an F adatom is adsorbed in the resulting vacancy, while in the ``Ag vacancy" case the vacancy in the surface already existed.} \label{Fig:Eads}
\end{figure}

The adsorption energies obtained for the Ag(100) surface are plotted in Fig.~\ref{Fig:Eads}\,(a). According to Eq.~\eqref{eq:eda2edm}, the reaction with atomic fluorine is more exothermic than the reaction with molecular fluorine. 

For the studied range of coverages, the most energetically favorable position is the hollow site, followed by the bridge, and finally, with a notable energy penalty, the top site. The smallest energy difference between the hollow and bridge sites is 140 meV, found at the two lowest coverages studied ($\Theta=1/18$ and 0.125 ML). This result indicates that the hollow site is the preferred choice for F adatoms when the Ag(100) surface is fluorinated. The hollow sites also show the smallest increase in adsorption energy with $\Theta$, indicating weak F-F interactions.

We can use the above result to estimate the surface diffusion constant. To jump from one hollow site to another the adatom should climb to the bridge site which determines the activation barrier $\Delta E_{ad}=$140~meV corresponding to 1600 K. Although this is larger than room temperature we can estimate a jump hopping rate as $\Gamma=\nu \exp(-\Delta E_{ad}/k_B T)=10^{10} $~jumps/sec where, as an order of magnitude,  we estimated $\nu=$15~THz by a typical Ag-F stretching phonon frequency~\cite{Gawraczynski2019} in \ch{AgF2}. This indicates that at room temperature there is substantial diffusion and the system can equilibrate at least at the length scale of the STM experiment. At 53 K the jumps become of the order of one per second and, at the typical temperature of the STM experiment (10~K), the adatoms are essentially frozen. 

Because F is the most electronegative element charge transfer occurs from the Ag to the F atom (Table~\ref{Table:atom_dist}). Since the F is above the surface this creates a dipole moment. Therefore at small coverages, the interaction energy among adatoms will be dominated by dipole-dipole interactions. This yields a contribution to the adsorption energy behaving as
\begin{equation}\label{eq:dip}
    \delta E_{ad}=\alpha \frac{p^2 \Theta^{3/2}}{(a')^3},
\end{equation}
where $\alpha$ is a constant of order 1 which depends on the Wigner-like lattice the dipoles form and $p$ is the dipole moment. The order of magnitude can be obtained by setting $\Theta=0.5$ ML and $\alpha=2$, which corresponds to truncating the dipole interaction up to the nearest neighbor in a square lattice. If one assumes a full electron is transferred to the F one obtains $\delta E_{ad}=1.8$~eV. This is larger than the adsorption energy change at small coverage. However, charge transfer is smaller than one electron which is reflected by the small $\delta n$ in Table~\ref{Table:atom_dist}. Using the charge on the sphere to compute the dipole moment gives a value that is too small as it underestimates the dipole moment. A more accurate computation of the dipole moment is beyond our scope, but one sees that a larger $d$ in Table~\ref{Table:atom_dist} for the top site (corresponding to a larger dipole moment) tends to yield larger interactions reflecting in a steeper $E_{ad}$ for the top site at small coverage. 

\subsubsection{Ag(110) surface}\label{sec:eadag110}

The adsorption energies for the Ag(110) surface are plotted in the 	Fig.~\ref{Fig:Eads}\,(b). For the lowest coverage studied, the most favorable site is the short bridge. The long bridge and hollow  sites have an energy penalty $\Delta E_{ad}^L=90$~meV and  $\Delta E_{ad}^H=140$~meV  respectively, similar to the energy penalty of the bridge site in the (100) surface.  
 
The top site has the largest $d$ in Table~\ref{Table:atom_dist} and the largest $\Theta$ dependence for $\Theta\leq 0.5$ ML. This is again consistent with dipole-dipole interactions in Eq.~\eqref{eq:dip}. Among the lowest energy sites, the short bridge has the largest $d$ which explains why its energy converges to the other sites having much smaller dipole moments and negligible dipole-dipole interactions. The energy differences reach a minimum at $\Theta=0.5$ ML where the three sites become almost degenerate in agreement with Wang et al~\cite{Wang2001}.

Absolute results for $E^{(a)}_{ad}$ at $\Theta=0.5$ ML differ by $\approx8\%$ from those of Wang (considering that they use the opposite convention for the sign of the energy). In the computation of Eq.~\eqref{eq:mu0a}, we used the ground state of the F atom, which happens to be spin-polarized. If we instead restrict the atomic computation to an unpolarized state, we find $E_d=1.90$~eV, and the agreement between our adsorption energies and Wang's is within 2\% for all the available cases.

Since to escape from a short bridge the adatom has to pass through the hollow site, the lowest diffusion barrier at low coverage will be given by $\Delta E_{ad}^H$ ($\approx$~1600 K), which coincides with the Ag(100) surface. 
However, to move to the nearest neighbor short bridge in the horizontal direction in Fig.~\ref{Fig:surf_param} requires the following passages: short bridge$\rightarrow$ hollow $\rightarrow$ long bridge $\rightarrow$ hollow $\rightarrow$ short bridge. In contrast, the movement in the perpendicular direction requires only one passage through a hollow site. This can make diffusion below 1600 K considerably anisotropic due to different prefactors of the activated (exponential) temperature dependence. 

Increasing the coverage, as soon as nearest neighbor short-bridge sites become occupied
($\Theta>0.5$ ML) the energy of the short-bridge location increases steadily with respect to the others. This indicates that the short F-Ag-F motive has a considerable energy penalty with respect to a long  F-Ag-F motive. Notice that these changes occur at coverages much larger than the ones studied in Ref.~\cite{Sanchez2024} ($\Theta< 0.02$ ML). 

The hollow and long-bridge adsorption energies remain close to each other for the whole range of coverages studied and they increase weakly with coverage,  indicating weak interactions for these positions. 

The adsorption energy of the top site for Ag(110) is, once again, the less favorable one out of the four possible sites for every value of the coverage studied.

\subsubsection{Exchange reactions and vacancy occupancy}
\label{sec:exchreac}

We also considered the possibility that fluorine adatoms substitute surface Ag atoms, which are incorporated into the bulk. 
A viable path for such an exchange reaction process is that an Ag atom from the surface migrates to a kink along a step of the surface~\cite{Scheffler2000}. Since this only moves the kink by one lattice constant it is easy to see that the energy of the silver increases by $\mu^0_{\ch{Ag}}$ as dictated by Eq.~\eqref{eq:em2}. 

Figure~\ref{Fig:Eads} shows the exchange reaction adsorption energy for the two surfaces at low coverages (green diamonds). From an energetic point of view, these reactions have a penalty that is similar to the top site although somewhat less favorable. 

Another possibility is that the fluorine adatom occupies a vacancy site that is already present, saving the energy cost to expel the Ag. The energy in this case is given by the orange hexagons in Fig.~\ref{Fig:Eads} and it is still less favorable than the adsorption of the adatom on one of the highly-symmetric sites of the ideal surface, with the exception of the top site. 


\begin{figure}[t]
  \centering
  \includegraphics[width=1.1\columnwidth]{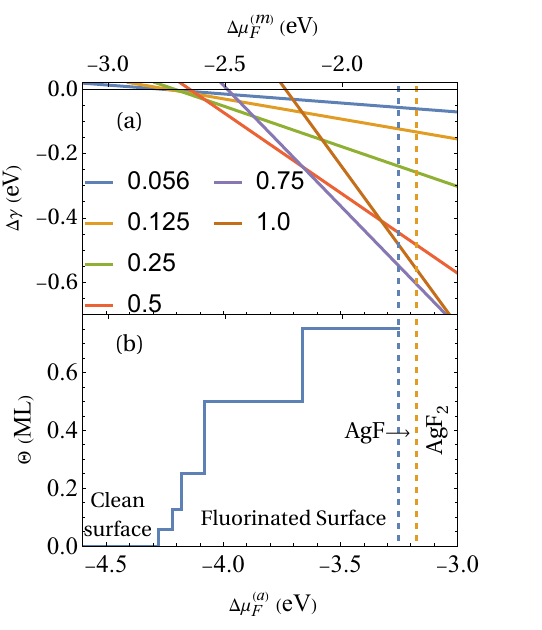}
  \caption{
  (a) Change of surface Gibbs free energy as a function of the atomic (lower scale) and molecular (upper scale) chemical potential for the Ag(100) surface. Labels indicate the coverage of each line. (b) Most stable surface coverage vs. fluorine chemical potential. The lower (upper) scale corresponds to the atomic (molecular) chemical potential. The dashed line indicates the limit of stability of the bulk phases: AgF between the dashed lines and \ch{AgF2} to the right of the orange dashed line. 
    \label{fig:gdm}}
\end{figure}

\subsection{Thermodynamics}\label{sec:thermoresults}


We use the formalism of Sec.~\ref{sec:thermo} to estimate the thermodynamic stability of the different coverages. We focus on the Ag(100) surface but a similar analysis can be done for the Ag(110) surface using the reported values of adsorption energies. Since the pressure and temperature dependence is through the chemical potential it is enough to obtain the minimum Gibbs free energy as a function of $\Delta \mu_F$ [Eqs.~\eqref{eq:dgdtp},~\eqref{eq:dgdtp0}]. 

Figure~\ref{fig:gdm}\,(a) shows the change in the surface Gibbs free energy as a function of chemical potential for various coverages and both gases (upper and lower scales). $\Delta \gamma<0$ corresponds to the fluorinated surface being more stable than the clean surface. Similar computations for the oxide are shown in Fig. 7 of Ref.~\cite{Li2003a} where the $x$ axis label should read $\Delta \mu_O$. The stable structure is given by the line segment with the lowest surface Gibbs free energy. Panel (b) shows the evolution of the surface coverage as a function of the chemical potential. Plateaus indicate that the corresponding structures will be stable in a certain chemical potential range. The silver surface is in a clean state for a very negative chemical potential. Increasing the chemical potential a progressively larger fluorine coverage becomes stable. Notice that the plateau is particularly extensive for $\Theta=0.5$~ML hinting at a relatively large range of pressure and temperature where this phase is stabilized. 

\begin{figure}[t]
  \centering
  \includegraphics[width=1\columnwidth]{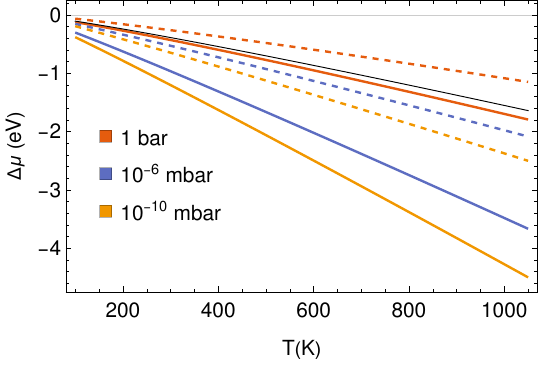}
  \caption{Temperature dependence of the chemical potential for different pressures using experimental data from the National Institute of Standards and Technology~\cite{Linstrom2023}. The full (dashed) lines correspond to the monoatomic (diatomic) fluorine gas. The thin line is the ideal monoatomic gas result (Eq.~\ref{eq:ig}) for a pressure of 1 bar. 
    \label{fig:dmudt}}
\end{figure}

The dashed lines indicate the range of thermodynamic stability of the bulk phases. These were obtained as before using zero temperature energies to estimate bulk Gibbs free energies~\cite{Reuter2001}. The fluorine chemical potential at which the reaction, $\ch{Ag +  F <=> AgF}$ is in thermodynamic equilibrium (dashed blue line) satisfy,
\begin{equation}
      \Delta  \mu_{F(\ch{AgF})}^{min}(T,p)=
 E_{\ch{AgF}}^{bulk}-E_{\ch{Ag}}^{bulk}-\mu^0_F,
     \label{eq:muminAgF}
\end{equation}
where the energies are per formula unit. Below this value, AgF decomposes by losing fluorine. Analogously the lower limit of stability of \ch{AgF_2} (orange line) is determined by  
\begin{equation}
    \label{eq:muminAgF2}
  \Delta  \mu_{F(\ch{AgF2})}^{min}(T,p)=
 E_{\ch{AgF2}}^{bulk}-E_{\ch{AgF}}^{bulk}-\mu^0_F.
\end{equation}
    
Figure~\ref{fig:dmudt} shows the chemical potential as a function of temperature for a pressure of 1 bar ($\approx0.99$~atm) and typical pressures in ultrahigh vacuum experiments. At room temperature (300 K) for all reported pressures and gases the thermodynamic equilibrium is given by bulk fluorinated phases. This means that in fluorination experiments as in Ref.~\cite{Sanchez2024} the state of the surface is not dictated by thermodynamics but by kinetics factors as the time of exposition and the sticking coefficient. Even if the surface were exposed to the gas for a very long time, bulk compounds would not be formed due to the appearance of Cabrera-Mott barriers blocking the reaction~\cite{Cabrera1949,Qiu2001}. 

\begin{figure*}[t]
    \begin{center}
        \includegraphics[width=1\textwidth]{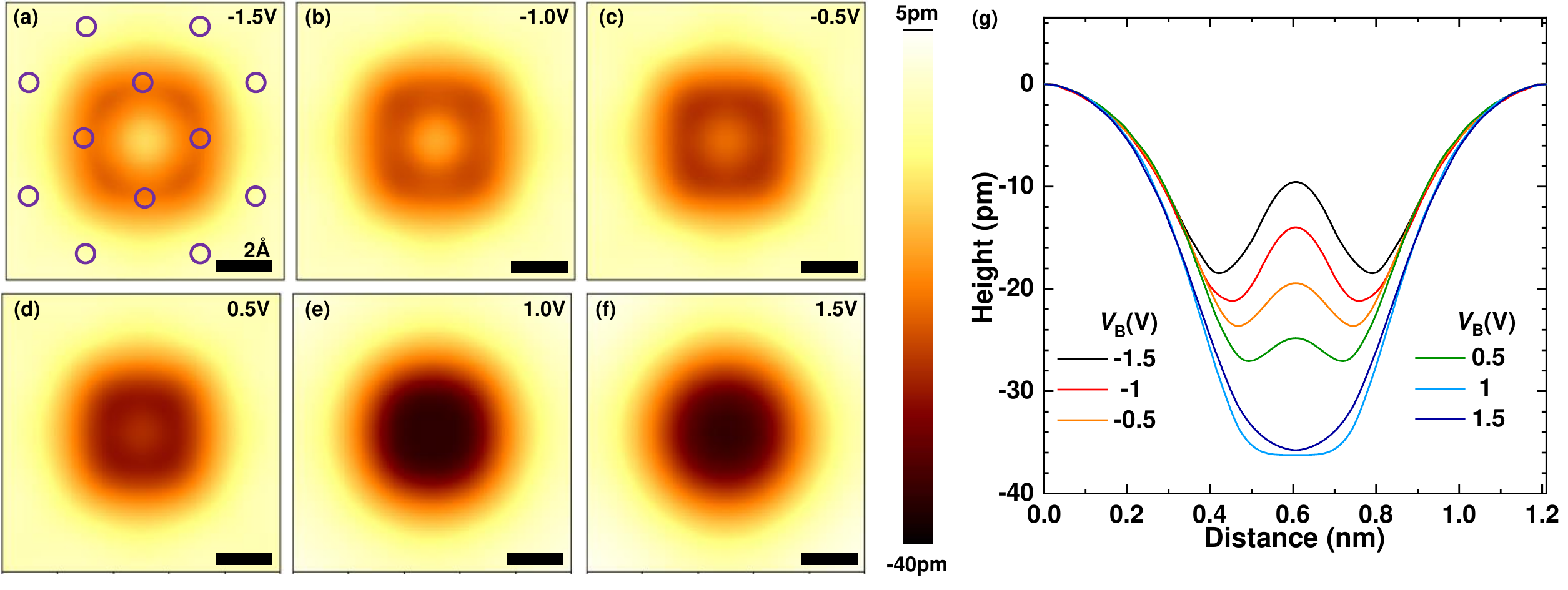}
    \end{center}
    \caption{(a)-(f) Simulated STM topographies using DFT and the Tersoff-Hamann approximation for a F adatom in the hollow position of the Ag(100) surface. The values of the bias voltage range from -1.5 to 1.5 V with a tunneling current of $5\times 10^{-7}$ a.u., corresponding to 3.3 nA. The superimposed purple circles on panel (a) correspond to the positions of silver atoms on the surface of the slab. The scale bar corresponds to 2 \AA{} in each panel. (g) Height profiles of F atoms for the studied bias values. The profiles were taken parallel to the horizontal axis of the topographies, crossing over the F adatom, with the reference value for the apparent height of the tip taken at the leftmost extreme of the profile curve. These reference values range from 3 to 5 \AA{}, which compares well with typical distances for tunneling conditions. See  Fig.~3 of the SI~\cite{Note2} for actual values.} \label{Fig:100STM_prof}
\end{figure*}

From Fig.~\ref{fig:gdm} we see that a clean surface ($\Delta \gamma>0$) requires a monoatomic chemical potential more negative than -4.3 eV which, according to Figure~\ref{fig:dmudt} requires temperatures larger than 1000 K and minimal pressures. In contrast, the reaction with diatomic fluorine is stable at $T=1000$~K and low pressure. Thus, this computation predicts that the surface will act as a molecule splitter at high temperatures. Molecular fluorine will be absorbed and atomic fluorine will be released.

\subsection{Scanning tunneling topographies}\label{sec:topographies}
\subsubsection{Fluorine adatoms on the Ag(100) surface}

The adsorption energy calculations predict the hollow site as the most favorable position of a F atom adsorbed on Ag(100). Unfortunately, atomic resolution requires the tip to be very close to the sample, which perturbs the adatom position. Ref.~\cite{Sanchez2024} circumvented this problem by a ``split-image" method. Two contiguous regions were measured with different tip heights and the images were matched keeping them in register. For the Ag(100) surface, they report the adatoms in the hollow site in agreement with our result as this being the more favorable site. 

To further verify this result we computed the STM AT for this case (Fig.~\ref{Fig:100STM_prof}). A clear sombrero-shaped AT is observed for negative bias with the apex of the sombrero suppressed as the bias increases. Comparing these results with those reported in Fig. 2 of   Ref.~\cite{Sanchez2024}, we observe that this behavior aligns with experimental observations. 

At $V_{\rm B}=1.0$~V the center of the feature reaches its minimum and the sombrero becomes a depression. With further increase in bias voltage (beyond the range studied experimentally in Ref.~\cite{Sanchez2024}), the depth of the depression decreases [see Supplementary Information (SI)~\footnote{See Supplemental Material at [URL will be inserted by publisher] for all the ATs computed and the corresponding reference values of the apparent height of the tip.} Fig.~1].

We do not expect theoretical bias voltages to match the experimental ones exactly. Kohn-Sham~\cite{Kohn1965} DFT maps the system to a non-interacting problem that has the same electronic density as the interacting one. In contrast to the density, the Kohn-Sham energies are auxiliary quantities without a priori correspondence to the binding energy of interacting quasiparticles. Typically, due to interactions, experimental quasiparticle bands are narrower than Kohn-Sham bands. Therefore, we expect bias voltage magnitudes to be moderately overestimated in DFT. This may explain why for negative bias the protrusion appears less developed in the theory than in the experiment (a more negative theoretical bias is needed to match the experiment). For $V_{\rm B}=1.0$~V, both theory and experiment show a pure depression, suggesting this discrepancy is smaller.   

While the behavior of the experimental STM ATs as a function of the bias voltage is well reproduced, we observe that the depth of the feature is overestimated, and its width is underestimated by approximately a factor of 0.5. This discrepancy can be attributed to the inherent simplifications in the Tersoff-Hamann approximation: {\it i}) the tip is modeled as a spherical $s$-wave, which may not accurately represent the tungsten tip used in Ref.~\cite{Sanchez2024}; {\it ii}) states are assumed to be unaffected by the tip potential, whereas, in reality, the electronic cloud around fluorine can polarize substantially depending on the tip polarity~\cite{Schott1994}; {\it iii}) DFT maps the interacting problem to a non-interacting one, and using non-interacting wave functions and eigenvalues to model the tunneling conductance can be a crude approximation; {\it iv}) lateral resolution issues can also affect the apparent width and depth of the features.

We also computed the STM images for the adsorbate in the bridge position (not shown). In this case, we obtained a sombrero shape across the entire range of experimental current and bias voltage. This contrasts with the experimental findings, which show a depression AT for positive bias. Specifically, the depression was observed only for bias values larger than 0.5 V and tunneling currents lower than 662 pA. These results further confirm that the observed AT corresponds to the hollow site. 

\begin{figure*}
    \begin{center}
\includegraphics[width=1\textwidth]{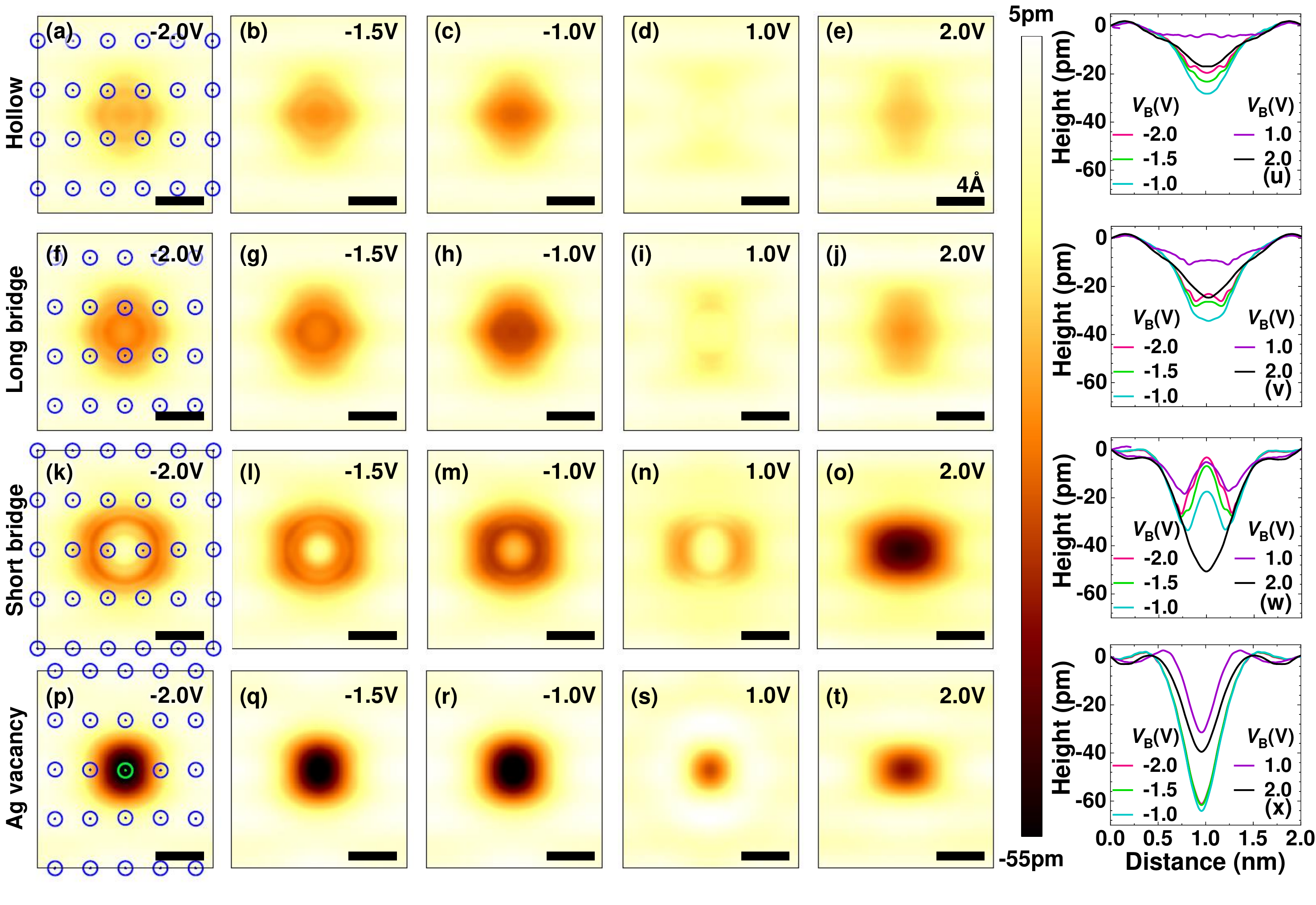}
    \end{center}
    \caption{STM topographies simulated using DFT and the Tersoff-Hamann approximation for the Ag(110) surface for an F adatom in the (a)-(e) hollow, (f)-(j) long-bridge, (k)-(o) short-bridge positions, and (p)-(t) F adatom on a surface Ag vacancy. The scale bar corresponds to 4 \AA{} in each panel.
   The bias voltage values range from -2 to 2 V with a current of $1\times10^{-8}$ a.u., corresponding to 66 pA.
   Blue circles overlaid on panels (a), (f), (k) and (p) indicate the positions of the silver atoms on the slab surface while the green circle in (p) indicates the substituted atom. The right-hand side panels display the corresponding profiles taken along diagonal lines with equal horizontal and vertical coordinates ($x=y$), passing through the fluorine position.
 We set the zero value for the apparent height of the microscope tip at the point in this line with the largest separation from the F adatom and its image in the neighboring supercells. The absolute value of the height for each voltage is reported in Figs.~6, 9, 12 and 18 of the SI~\cite{Note2}
   These reference values range from 5 to 6 \AA{}, which compares well with typical distances for tunneling conditions. } \label{Fig:110STM_prof}
\end{figure*}

\subsubsection{Fluorine adatoms on the Ag(110) surface}\label{sec:at110}

In the experiments of  Ref.~\cite{Sanchez2024}, three different ATs were identified on the Ag(110) surface: AT(A), the most frequent one, showing the deepest depression; AT(B), with intermediate frequency and depression depth; and AT(C), the rarest with the smallest depression depth. The reported abundance is AT(A) 60\%, AT(B) 35\%, and AT(C) 5\%. Assuming as an order of magnitude that the experimental frequencies reflect thermal populations at room temperature we obtain adsorption energy differences $\lesssim 60$~meV, of the same order of the $\Delta E_{ad}\sim$ 90-140~meV reported in Sec.~\ref{sec:eadag110}. 

It is natural to assign the experimental ATs to the three more stable locations in DFT which are also close in energy. Figure~\ref{Fig:110STM_prof} shows the simulated ATs (first three rows). Comparing the profiles in the experiment and the theory (right panels in Fig.~\ref{Fig:110STM_prof}) and using the depth of the wells as a criteria AT(A) would be attributed to the short bridge, AT(B) to the long bridge and AT(C) to the hollow site. Assuming the lowest energy adsorbates would show the higher frequency leads to the same assignment. 

In Ref.~\cite{Sanchez2024}, the split-image method was used to determine the adatom position also for the Ag(110) surface. They report AT(A) to correspond to the short bridge as expected. However, the second more abundant location, AT(B) was assigned to the hollow site, and the relatively more favorable long bridge was not reported. Instead, AT(C) was assigned  to the top site which in our computations has significantly more energy [c.f. Fig.~\ref{Fig:Eads}\,(b)]. 

More insight can be obtained by comparing the topographies. The simulated AT for voltage biases ranging from $-2$ to $2$ V and a tunneling current of 66 pA are presented on the left side of Fig.~\ref{Fig:110STM_prof}. 
While all sites exhibit a depression for all voltages shown, in the case of the long and short bridges, a protrusion appears at the depression's center for negative bias giving the sombrero shape.

Among the three energetically more favorable locations, the short bridge has the deepest depression which is in agreement with the above assignment to the AT(A) having the deepest depression in the experiment. On the other hand, the protrusion is not present in the experimental AT(A). For the hollow site, the well has a smaller depression in partial agreement with the experimental AT(B). Regarding the protrusion, the situation is reversed: it appears in the experimental AT(B) but not in the theory for the hollow site.

The simplified orbital model presented in Sec.~\ref{sec:modelresults},  shows that the protrusion is very sensitive to the competing role of  F $2p$ and Ag $5s$ states. Its presence or absence depends on factors such as the relative strength of the symmetry-projected DOS in these channels, and the difference in decay length between F $2p$ and Ag $5s$ orbitals near the Fermi level. These aspects are susceptible to errors in DFT calculations. In contrast, the depth of the depression is mainly determined by the oxidation state of the Ag atoms surrounding the central F atom. This feature is more robust in DFT given that the density plays the central role in this technique. Therefore, we find the assignment of the AT(A) and AT(B) to the short bridge and the hollow site, respectively, as compatible with the theory. The depth of the depression (excluding the protuberance) is overestimated, which we attribute again to the simplifications of the Tersoff-Hamann approximation and lateral resolution effects.

It remains to assign the less abundant AT(C). A plain top position would require a topography with a large protuberance for positive bias. For example, for a bias voltage $V_{\rm B}=-2$ V and current as in Fig.~\ref{Fig:110STM_prof}, our computations predict a protuberance of 40~pm above the silver level which is incompatible with the experimental observation of depressions for all ATs. This together with the high energy cost [Fig.~\ref{Fig:Eads}\,(b)] allows to exclude a simple top position. 


Another possibility is that AT(C) corresponds to an exchange reaction (Sec.~\ref{sec:exchreac}). The energy penalty is even larger than for a top position [Fig.~\ref{Fig:Eads} (b)]. A third possibility is that F atoms fill already present vacant sites. Since the number of these sites is a small fraction of a small coverage, a tiny number of vacant sites would be enough to explain the observed abundance (of the order of 1/5000 fraction of vacant surface sites). However, filling these sites is still not more favorable energetically than the three previous sites [see orange hexagon in Fig.~\ref{Fig:Eads} (b)]. 

Either if the vacancy was already present or not the final configuration is the same. The last row in Fig.~\ref{Fig:110STM_prof} shows the ATs for this final configuration. Interestingly, the experimental AT shows a rim around the depression (volcano shape) for positive voltage which is well reproduced by the theory [panels (s),(t), and (x)]. There are also other striking similarities between theory and experiment: for negative bias voltage, the depth of the ATs remains without significant changes while for positive bias the well rises rigidly enhancing the rim of the volcano [Fig.~4,(g) of Ref.~\cite{Sanchez2024}]. 

Unfortunately, the experimental AT(C) corresponds to the smallest depression while the theoretical one is the deepest.  This is due to the depression being mainly determined by the silver $5s$ states (see Sec.~\ref{sec:modelresults}) which here are retracted due to the vacancy.

\begin{figure}
    \centering
\includegraphics[width=\linewidth]{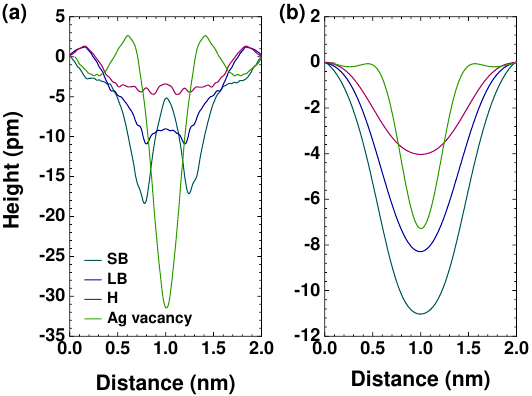}
    \caption{(a) Height profiles for a bias voltage of 1.0 V and the four studied locations: short bridge (SB); long bridge (LB); hollow (H) and Ag vacancy, taken as in Fig.~\ref{Fig:110STM_prof}.  (b) The same profiles after convoluting with a two-dimensional Gaussian with a FWHM of 4.7~\AA.}
    \label{fig:Lateral_resolution}
\end{figure}

Adding lateral resolution effects (which may also simulate a non-spherical tip) improves the agreement. As an example, Fig.~\ref{fig:Lateral_resolution} compares the height profiles as resulting form the computations (a) and convoluting with a two-dimensional Gaussian with full with at half maximum (FWHM) of 4.7~\AA{} (b). Several features improve the agreement with the experiment: i) Wells become shallower and wider. ii) The vacancy location (green) has a smaller well than the short bridge. iii) The short bridge loses the sombrero feature. Similar improvements are obtained for the Ag(100) case without fully losing the sombrero feature (not shown). On the other hand, for bias voltage -1.5 V (not shown), the vacancy still shows the deepest AT and the short bridge keeps the sombrero feature which contradicts the experiment. This problem casts doubts on the assignment of AT(C) to a fluorine substitution.  Furthermore, it remains unexplained why the long bridge is not observed despite being energetically more favorable than the hollow site.  Another possibility is that additional adatoms as hydrogen are playing a role. 
More experimental and theoretical work is needed to resolve this issue. 






\subsection{\label{sec:modelresults}Oxidation states and orbital origin of apparent topography}

To explain the transition between depression and sombrero shapes observed in the STM topographies, it is necessary to understand how different orbitals contribute to the tunneling current. Here we use the formalism of Sec.~\ref{sec:orbmodel} to link the appearance of the depression and sombrero shape to specific orbitals. For the case of the F adatom on the Ag(100) surface, the Wannier orbitals can be taken as transforming like cubic harmonics, i.e., $p_x$, $p_y$, $p_z$ and the absence of mixing among onsite orbitals in Eq.~\eqref{eq:approxdiag} becomes exact. In less symmetric situations, like the orbitals of an Ag next to an F adatom, Eq.~\eqref{eq:approxdiag} is a good approximation, as the onsite mixing of $5s$ and $4d$ orbitals will be very small. 

\begin{figure}
    \begin{center}
        \includegraphics[width=\columnwidth]{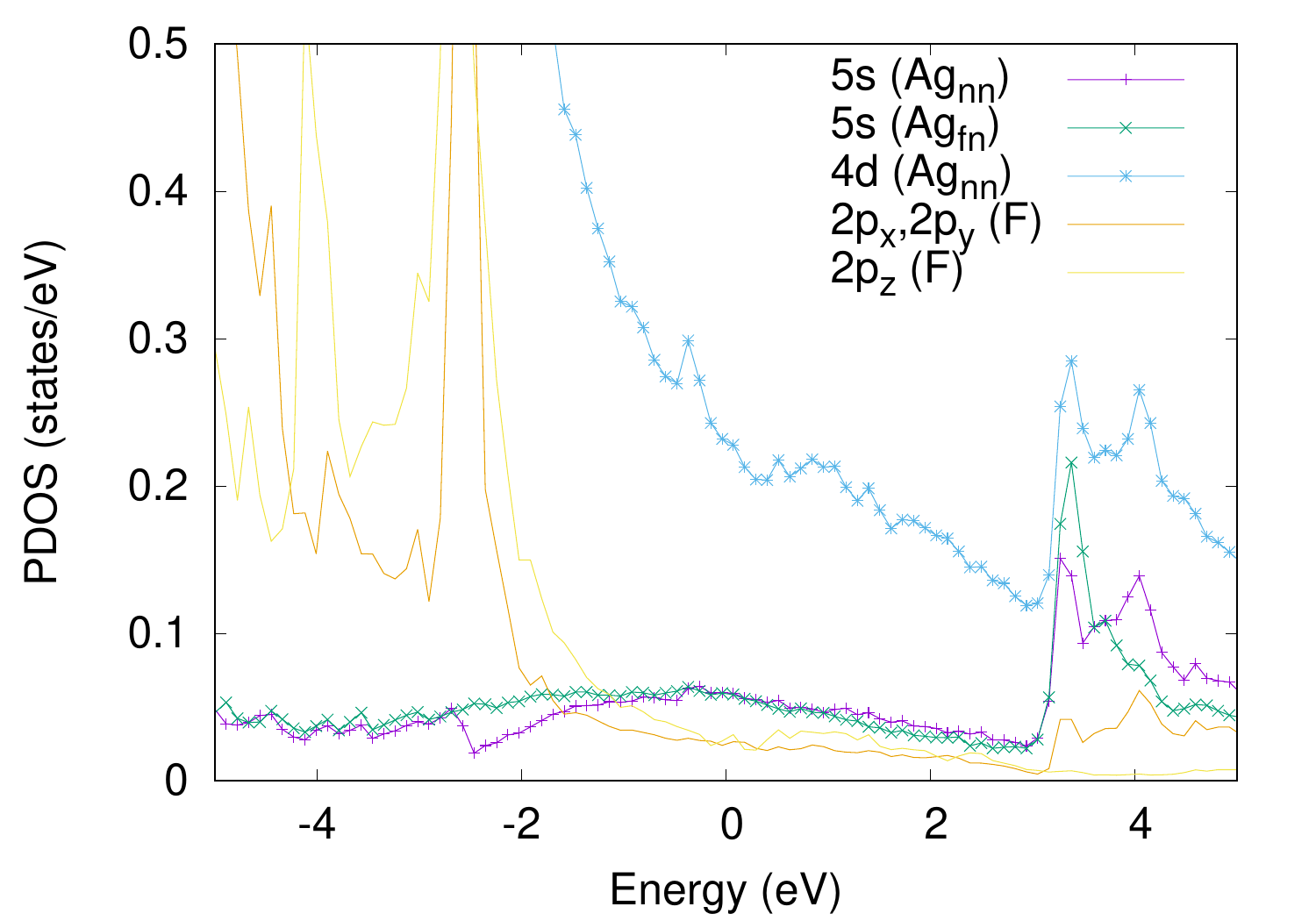}
    \end{center}
    \caption{PDOS of the F adatom located at the hollow site and its nearest neighbor (nn) silver as a function of the energy for the Ag(100) surface.
    For the silver $s$ states we also show the PDOS in the farthest neighbor (fn). 
    Computations were done in the $3\sqrt2\times3\sqrt2$ supercell ($\Theta=1/18$ ML).
    The zero value of the energy corresponds to the Fermi energy.
    The principal quantum number refers to the prevalent contribution in the present energy window.
    The PDOS for the different sites in the Ag(110) surface are qualitatively similar and are not shown.} \label{Fig:PDOS100_orbitals}
\end{figure}

As the microscope tip scans the surface of the system, the typical distance to the atoms is around $\sim 5-6$~\AA. Outside the slab, the atom-centered Wannier orbitals decay exponentially, indicating that at these distances, the tip interacts with the evanescent part of the wave function. Equations~\eqref{eq:bardeen},~\eqref{eq:Rxdos} show that significant tunneling currents require orbitals that not only have a large contribution to the PDOS but also exhibit a large spatial extension. 

\begin{figure}[tb]
	\begin{center}
        \includegraphics[width=1\columnwidth]{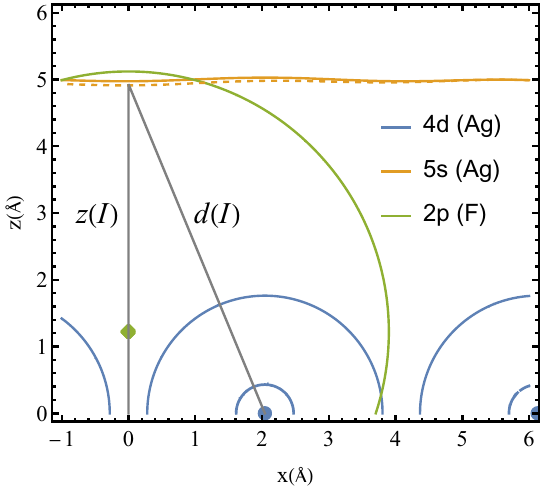}
	\end{center}
	\caption{Constant density plot of the contribution to the density from hydrogenic orbitals of different symmetry. 
   The $z$-coordinate is perpendicular to the Ag(100) surface plane while the $x$ coordinate runs along the [100] direction and intersects the Ag positions.
   The $y$ coordinate is perpendicular to the plot's plane.  
   $z=0$ corresponds to the topmost layer.
   The blue dots represent Ag atoms at $z=y=0$ and $x=\pm a/2,3a/2$.
   The other 4 Ag atoms contribute to the density and are placed at $z=0,x=0,y=\pm a/2$ and  $z=0,x=a,y=\pm a/2$.
  The green diamond is the fluorine in the hollow position.
  We represent the density of Ag orbitals as the sum of the densities of the individual atoms.
  For each orbital type curves correspond to a constant density $\rho_0=10^{-4}~\text{\AA}^{-3}$.
  The gray line shows the definition of tip-nearest neighbor Ag distance $d(I)$ and the height $z(I)$ in the case in which $I$ is determined by $\rho_0$ in the pristine surface.
  The dashed curve is the result in the case in which the planar Ag nearest neighbor to the F loses 1/4 of an electron.} \label{Fig:hydrogenic2d}
\end{figure}

We identify $D_{\bm R l}(\epsilon)$ with the partial-DOS in the silver slab obtained with VASP (Fig.~\ref{Fig:PDOS100_orbitals}). The zero value of the energy corresponds to the Fermi level of the system and, in the Tersoff-Hamann approximation, to zero bias voltage. The largest contribution to the PDOS in the energy range of interest (-2 to 2~eV) comes from the silver $d$-orbitals. However, as we will see, these have a too short range to contribute significantly to the current. Due to their large bandwidth, $W$, $s$-orbitals exhibit a smaller PDOS ($\sim 1/W$). At negative bias, there is a significant contribution from the F $2p$ states, a consequence of the $p$ shell of fluorine being full. 

\begin{figure}
    \begin{center}
        \includegraphics[width=1.\columnwidth]{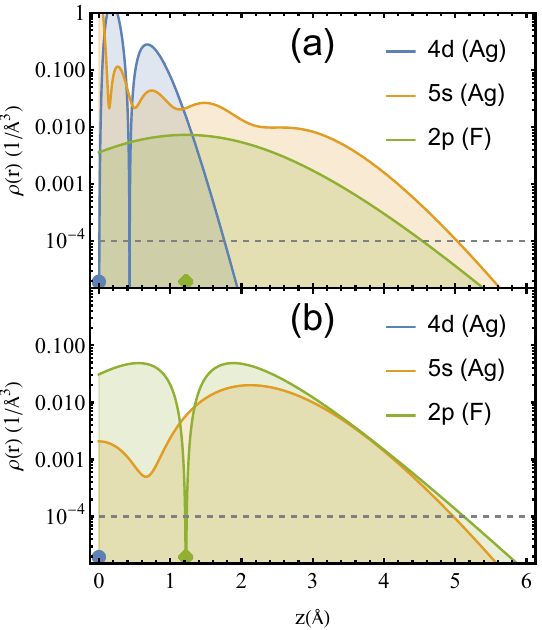}
    \end{center}
    \caption{Plot of the density of Fig.~\ref{Fig:hydrogenic2d} along the line $x=a/2$ (a) and $x=0$ (b).
    Blue dots and green diamonds represent the height of Ag and F atoms respectively.
    The $x$ axis is perpendicular to the picture's plane.
    Thus Ag is above and below the picture's plane in (a) and $F$ is underneath the picture's plane in (b).
    The horizontal dashed line corresponds to the constant density contours of Fig.~\ref{Fig:hydrogenic2d}.} \label{Fig:hydrogenic}
\end{figure}

As a proxy of the Wannier orbitals, we will use simple hydrogenic orbitals. We estimate the spatial extent of the orbitals using a screened effective charge $Z_{\rm eff}$ as proposed in the classical work of Slater~\cite{Slater1930} and computed in Ref.~\cite{Clementi1963}. This defines the Slater radius, $r_{S}=a_0n^2/Z_{\rm eff}$, given by the maximum of Slater's radial charge density, where $a_0$ is the Bohr radius and $n$ is the principal quantum number as in Ref.~\cite{Clementi1963}. Using this approach, we determine the values of the Slater radii to be $r_{S}^{4d}=0.57$~\AA~ ($Z^{4d}_{\rm eff}=14.763$) and $r_{S}^{5s}=1.96$~\AA~ ($Z^{5s}_{\rm eff}=6.756$) for Ag $4d$ and $5s$ orbitals, respectively. The values in parenthesis are the screened charges from Ref.~\cite{Clementi1963}. Despite the simplicity of the approximations, these radii compare reasonably well with Hartree-Fock (HF) values~\cite{Mann1968}, which are $r_{HF}^{4d}=0.54$~\AA~ and  $r_{HF}^{5s}=1.53$~\AA.

Assuming the adsorbed fluorine is in the $2p^6$ configuration (\ch{F^-}) the effective charge is not known. Therefore, we use the insight from Ref.~\cite{Slater1964} that the Slater radius for the outermost orbital is approximately equal to the ionic radius. We estimate  the effective charge as $Z_{\rm eff}^{2p}=a_0n^2/r_{\rm exp}=1.6$, with $n=2$ and $r_{\rm exp}=1.33$~\AA~ (Ref.~\cite{Shannon1976}). These estimates show that the Ag $4d$ orbitals are more localized than the $5s$ and F $2p$ orbitals, making them less likely to participate in the tunneling. Although the F $2p$ orbitals are shorter range than the Ag $5s$ orbitals,  as we shall see, this can be compensated by the fact that F is positioned higher than the topmost Ag layer.

We first consider the translational invariance case without the F adatom. In this case, the PDOS at the different silver atoms is identical, allowing us to drop the $\bm R$ label and write Eq.~\eqref{eq:Rxdos} as
\begin{equation}
    N(\bm r ,\epsilon)\approx\sum_{ l} D_{ l} (\epsilon) \rho_{ l}(\bm r),\nonumber
\end{equation}
with 
\begin{equation}
    \rho_{ l}(\bm r)\equiv \sum_{\bm R}\rho_{\bm R l}(\bm r).\nonumber
\end{equation}
Figure~\ref{Fig:hydrogenic2d} shows a contour plot of $\rho_{ l}(\bm r)$ for silver $4d$ and $5s$ orbitals, fixing $\rho_{ l}(\bm r)=10^{-4}~\text{\AA}^{-3}$. We observe that the $4d$ orbitals are indeed very short-ranged. This is further illustrated in Fig.~\ref{Fig:hydrogenic} (a), which shows a log plot of the same density along $z$ for $x=a/2$, where the origin corresponds to the silver position (blue dot). Despite their large PDOS, this simple computation shows that $4d$ orbitals are completely irrelevant for the tunneling. It is not always appreciated that a large PDOS does not necessarily imply a large tunneling current, as matrix element effects (understood as all prefactors different from the integrated PDOS) can strongly alter the picture. 

Using that $D_{5s}\;\rho_{\bm R 5s}\gg D_{4d}\;\rho_{\bm R 4d}$ at the tip position ($\sim 5$\AA~$\gg r_S^{4d}$)  the local DOS at the tip can be simplified as,
\begin{equation}\label{eq:n5s}
    N(\bm r ,\epsilon)\approx D_{ 5s} (\epsilon) \rho_{ 5s}(\bm r).
\end{equation}

We now consider the fluorine adatom and its filled $2p$ orbital shell. In the presence of the adatom, contributions to the current can be present through the surface and the adsorbate~\cite{Sautet1996}. Indeed, neglecting the disturbance of the silver orbitals (full lines in Figs.~\ref{Fig:hydrogenic2d},~\ref{Fig:hydrogenic}), we see that at the tip-relevant distance (3-6~\AA), the F $2p$ and Ag $5s$ orbitals have similar densities hinting that both contributions can be important. Notice that, as anticipated, panel (b) of Fig.~\ref{Fig:hydrogenic} shows that immediately above the fluorine the shorter range of the $p$ orbital (green) is compensated by the fact that F is displaced towards larger $z$ (green diamond). 

For positive bias, F $2p$ orbitals have a small PDOS, and the current will be dominated by the $5s$ states (Fig.~\ref{Fig:PDOS100_orbitals}). 

\begin{figure}[tb]
	\begin{center}
        \includegraphics[width=1\columnwidth]{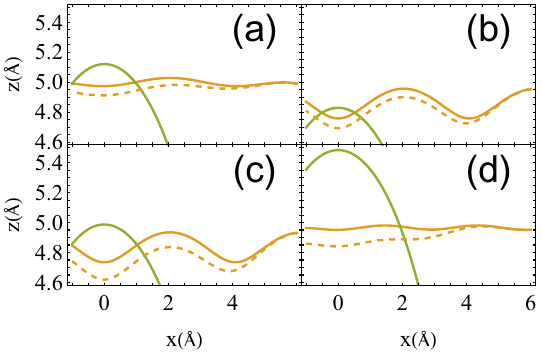}
    \end{center}
	\caption{(a) Constant density plot ($\rho=10^{-4}~\text{\AA}^{-3}$) of the contribution to the density from the $5s$ orbitals as in Fig.~\ref{Fig:hydrogenic2d} but in an expanded $z$ scale.
    The dashed (full) line corresponds to the oxidized (pristine) surface.
    Panel (a) corresponds to an F adatom adsorbed at the hollow site in the Ag(100) surface.
    The other panels correspond to the Ag(110) surface with the F adatom located at (b) the hollow, (c) the long-bridge, and (d) the short-bridge site.} \label{Fig:zdx4pn}
\end{figure}

First, we consider the Ag-tip (through surface) contribution. The effect of the F adatom is to modify the PDOS and the valence state of the silver neighbors due to oxidation. To consider the oxidation, we assume that $1/\zeta$ of an electron is transferred to the fluorine from the $\zeta$ nearest neighbor silvers surrounding it, where $\zeta=4$ is the planar coordination number for the hollow site. According to Slater rules~\cite{Slater1930}, this should decrease the screening of the core charges by 0.35 per electron. Thus, we increase the positive effective charge of these Ag atoms by $0.35/\zeta$. This leads to a contraction of the electronic charge distribution, as illustrated by the dashed line in Fig.~\ref{Fig:hydrogenic2d}. Neglecting possible changes in the PDOS, the depression of the charge density implies a depression of the tunneling current through Eqs.~\eqref{eq:bardeen},~\eqref{eq:Rxdos}. The zoom in Fig.~\ref{Fig:zdx4pn}\,(a) shows that the constant charge line retraction qualitatively mimics the depression observed in the STM experiments. This explains the tendency to have a depression-shaped feature for a positive bias in the measurements of Ref.~\cite{Sanchez2024}.

Panels (b)-(d) of Fig.~\ref{Fig:zdx4pn} correspond to the Ag(110) surface for the hollow, long-, and short-bridge positions, respectively. In all cases, the depth of the oxidation well is proportional to the charge depletion per Ag atom surrounding the F adatom. When the F adatom is in the hollow site it removes 1/4 of an electron from the surrounding Ag atoms ($\zeta=4$). This results in depression depths, roughly half of the depth of the case when the F atoms are in bridge positions and remove 1/2 of an electron from two F atoms ($\zeta=2$). In DFT we find that the charge transfer in the Ag $4d$ and Ag $5s$ orbitals have opposite signs so the $5s$ charge transfer is larger in magnitude than the total charge transfer in Table~\ref{Table:atom_dist}. The $5s$ charge transfer is -0.036, -0.033, -0.045, -0.05 for the Ag(100) hollow site and Ag(110)  hollow, long-bridge, and short-bridge sites respectively which is consistent with our arguments. For example, the $4s$ charge transfer is smaller in magnitude in the long-bridge than in the short-bridge site, corresponding to a shallower depression in Fig.~\ref{Fig:110STM_prof}.

\subsubsection{Analytical estimate of the depression depth}

It is instructive to compute analytically the factors that determine the depression depth in the present model. Restricting to the contribution of silver atoms, the current is written as,
\begin{equation}\label{eq:TH}
    I(\bm r,V_{\rm B})\approx\frac{4\pi e}\hbar|M|^{2} N_t(0)\sum_{\bm R}  G_{{\bm R}{5s}}(\epsilon)  \rho_{\bm R 5s}(\bf r),
\end{equation}
where we used the same approximations as in Eq.~\eqref{eq:n5s} and defined the cumulative DOS, 
$$G_{\bm R l}(V_{\rm B})=\int_0^{V_{\rm B}} D_{\bm R l}(\epsilon)d\epsilon.$$

For a given measurement current $I$, we define the depression's depth, $\Delta z=z_0-z_{\infty}$, as the difference between the tip height above the position of the fluorine taken as $\bm R=0$ (i.e. $x=y=0$ in Fig.~\ref{Fig:zdx4pn}) and the tip height in pristine site, $z_{\infty}$,  i.e. at the center of a hollow site far from the F. Here and in the following we use 0 and $\infty$ to label quantities at the F and pristine sites.  Furthermore, as we are interested in the depression depth, we turn off the contribution to the current through the F orbitals. As shown in the computation of Appendix~\ref{sec:app} the depression depth is determined by the 
constant height charge and cumulative DOS differences, namely,   
$\Delta \rho_{5s}\equiv\rho_{05s}(z_\infty)-\rho_{\infty5s}(z_\infty)$ and 
$\Delta G_{5s}\equiv G_{05s}-G_{\infty5s}$ 
and reads,
\begin{equation}\label{eq:dzdr}
    \Delta z=-\rho_{\infty 5s}\left(\frac{\partial  \rho_{{0} 5s}}{\partial z}\right)^{-1} \left(\frac{\Delta \rho_{5s}}{\rho_{{\infty} 5s}}+\frac{\Delta G_{5s}}{G_{{\infty} 5s}}\right).
\end{equation}
Remarkably, matrix elements, bias voltage, and coordination number factors cancel out. This equation shows that the reconfiguration of the nn silver wave functions and integrated PDOS determines the depression depth. From Fig.~\ref{Fig:PDOS100_orbitals} we see that for the bias voltage range of interest, the relative change of the integrated $5s$ PDOS (i.e. the area from the Fermi level to the bias voltage in $s$ symmetry) is practically unchanged when comparing the site nearest and furthest from the $F$ atom. Even at $V_{\rm B}=-2$~eV, ${\Delta G_{5s}}/{G_{{\infty} 5s}}\ll 1$. We therefore neglect this term and analyze the effect of the change in the density. In Eq.~\eqref{eq:rhonl} we provide the expression for the charge distribution for general hydrogenic orbitals. 

With the simplifications of Appendix~\ref{sec:app}, the change in the wall depth to leading order in $\Delta Z_{\rm eff}^{nl}$ and for general hydrogenic orbitals reads, 
\begin{equation}\label{eq:dzdz}
   \Delta z=-\left(\frac{d(I)}{Z_{\rm eff}^{nl}}-
\frac{3n a_0}{2 (Z_{\rm eff}^{nl})^2} \right) \frac{ d(I)}{ z(I)}\Delta Z_{\rm eff}^{nl},
\end{equation}
where $d(I)$ is the tip-nn Ag distance as shown schematically in Fig.~\ref{Fig:hydrogenic2d}. $\Delta z<0$ represents a depression corresponding to a transfer of $\delta/\zeta$ electron to a $F^{-\delta}$ ion. Within Slater rules $\Delta Z_{\rm eff}^{5s}=0.35 \delta/\zeta>0$. This simplified orbital model predicts that the depression depth: {\it i}) is independent of bias (assuming the bias does not affect significantly the distance, c.f. Fig.~3 of the SI~\cite{Note2}); {\it ii}) is proportional to the oxidation of the nn Ag atoms; and  {\it iii}) grows approximately linear with tip height ($d\approx z$, c.f. Fig.~\ref{Fig:hydrogenic2d}). With DFT, we find that {\it i}) is approximately fulfilled when the effect of the F protrusion is subtracted. For example, in Fig.~\ref{Fig:110STM_prof} (r) all curves overlap in the region away from the protrusion. A similar tendency is observed in the experiment, for example in  AT(A) shown in Fig.~4 (f) of Ref.~\cite{Sanchez2024} which is indeed independent of bias. Also, within DFT the bridge sites ($\zeta=2$) yield deeper walls than the hollow sites ($\zeta=4$) in agreement with {\it ii)}. 
Point {\it iii)} should be taken with a pinch of salt as it is valid away from the oscillatory region of the wave function but the $5s$ states have oscillations at quite large distances [c.f. Fig.~\ref{Fig:hydrogenic}\,(a)]. At shorter distances, the distance dependence will depend on the details of the wave function and requires the use of the full hydrogenic expression, instead of the asymptotic one leading to Eq.~\eqref{eq:rhonl}.

For the presented parameters, the modeled depression depth is smaller than the one obtained in DFT. This discrepancy can be attributed to the fact that the atomic orbitals have a decay length that is too short. Indeed, the decay length is determined by the Ag $5s$ ionization energy, which is significantly larger than the work function of silver. Adjusting $Z_{\rm eff}$ so that the ionization energy coincides with the work function of silver and using $Z_{\rm eff}=0.35 /\zeta$, and $z=6$ \AA, Eq.~\eqref{eq:dzdz} yields $|\Delta z|\approx 15,30$ pm for $\zeta=4$ (hollow) and $\zeta=2$ (bridge). These values are close to the DFT results. Unfortunately, for these parameters, the wave function widens and the tip falls in the region of wave function oscillations. An expression like Eq.~\ref{eq:dzdr} is still valid but, as already mentioned, the distance dependence requires detailed computations beyond our present scope.

\begin{figure}[tb]
	\begin{center}
        \includegraphics[width=1\columnwidth]{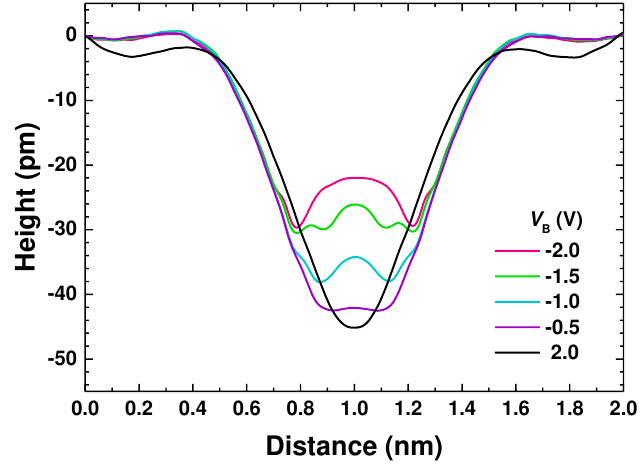}
    \end{center}
	\caption{Height profiles with the F adatom at the short-bridge position lowered 25 pm respect to the equilibrium position.
    The corresponding equilibrium computation is shown in Fig.~\ref{Fig:110STM_prof}(w).} \label{Fig:sb_lowerd_adatom}
\end{figure}

Regarding the sombrero shape, exactly above the fluorine atom, the constant density contour of the F $2p$ states protrudes beyond the surface defined by the Ag $5s$ contour. According to  Eq.~\eqref{eq:bardeen}, for sufficiently large negative bias where the F $2p$ PDOS dominates (Fig.~\ref{Fig:PDOS100_orbitals}), this protrusion of the F $2p$ density translates into a protrusion of the apparent constant current height at the center of the ``oxidation well" described earlier. Taking both effects into account leads to the following expression for the height of the protrusion relative to the bottom of the depression,
\begin{equation}\label{eq:dzfdr}
    \Delta z=-\left(\frac{G_{5s}(V_{\rm B})}{G_{2p}(V_{\rm B})}\zeta\frac{\partial  \rho_{{0}   5s}}{\partial z}+\frac{\partial  \rho_{2p}}{\partial z}\right)^{-1} \rho_{2p}(z),
\end{equation}
Here, the feature explicitly depends on the bias voltage, as found experimentally and in DFT computations. The strength of the protrusion increases as the bias voltage reaches regions with large F $2p$-PDOS. The protrusion becomes negligible when $|G_{2p}(V_{\rm B})|\ll |G_{5s}(V_{\rm B})|$ or when the density of the evanescent wave of the $2p$ orbitals diminishes significantly compared to that of the $5s$ orbitals. Therefore, the protrusion at different locations can appear when the densities are comparable at the tip distance, as illustrated in Fig.~\ref{Fig:hydrogenic}.

The green lines in Fig.~\ref{Fig:zdx4pn} illustrate how the relative height of the fluorine protrusion changes for the F in the different locations of the surfaces studied. In the simplified orbital model, this height depends only on the geometric arrangement of atoms and the range and local DOS of the involved orbitals. By adjusting the height of fluorine based on DFT results, we can explain the trends observed in the simulated STM topographies: the short bridge shows the largest protrusion (d), while the Ag(110) hollow case exhibits the smallest protrusion. Conversely, the Ag(100) hollow site and the long bridge show intermediate protrusions, consistent with  Figs.~\ref{Fig:100STM_prof} (g) and Figs.~\ref{Fig:110STM_prof} (p)-(r). Equations~\eqref{eq:dzdr} and \eqref{eq:dzfdr} provide qualitative trends for both the Ag depression and the strength of the F protrusion. 

As a practical check of the insight gained by the simplified orbital model, we recompute the DFT AT for the short-bridge site by artificially lowering the fluorine height by 25 pm (Fig.~\ref{Fig:sb_lowerd_adatom}). Comparing this with Fig.~\ref{Fig:110STM_prof} (w), it is evident that the maximum of the protrusion is suppressed by a similar amount, as expected from the model. As a byproduct, this suggests that lowering the position of the F can potentially reconcile the theoretical prediction of the short bridge with the experimental observation that shows no protrusion in the AT(A). Unfortunately, the displacement of the F is too large to attribute it to a DFT error on the structure. 

	\section{\label{sec:Conclusions}Summary and Conclusions}

In this work, we presented DFT computations of fluorine adatoms on Ag(100) and Ag(110) surfaces to identify the most energetically favorable adsorption sites.
The DFT energies were used to estimate the fluorine pressure and temperature regimes to stabilize the different phases. At typical conditions of fluorine experiments, thermodynamics predicts bulk fluorinated phases. This does not occur in practice because of kinetic factors such as a small sticking coefficient combined with a finite exposure time.  

The energy computations were complemented with simulations of the AT of fluorine in different high-symmetry locations
using the Tersoff-Hamann approximation. A simplified orbital model of the topography allowed us to interpret the DFT results.

The observation in Ref.~\cite{Sanchez2024} of only one kind of topography of adatoms in the Ag(100) surface was identified with the hollow site location being significantly more stable than the others.  This hypothesis was corroborated by simulating the STM topographies obtained from an F adatom situated on the hollow site, finding a good agreement for the trends as a function of  $V_{\rm B}$ with the experimental results. This also agreed with the experimentally found location using a split image method. 

For the Ag(110) surface, three different ATs were observed in the STM topographies of Ref.~\cite{Sanchez2024} for the F adsorbates labeled A, B, and C in order of depression depth and also of population.  As an order of magnitude, the population ratios are compatible with equilibrium Boltzmann factors at experimental temperatures and differences in energies of the order of  $\lesssim 60$~meV. 
DFT predicts energy differences of the same order for the three lowest energy locations.
This makes it tempting to assign these locations to the observed AT. Indeed, using the split image method, the short bridge, which is the more stable location, was attributed~\cite{Sanchez2024} to AT(A) and the hollow site to AT(B). However, the long bridge which is the second more stable location was not reported and instead AT(C) was reported to correspond to a top position. A fluorine adatom occupying a surface vacancy may explain the top site location. The AT has some similarities with the experimental one in that it presents a volcano shape and a similar bias dependence.  However, the well is substantially deeper than the experiment. 



Another important difference with the experiment is that the short bridge in DFT showed the switch from depression to sombrero, whereas the experiment for AT(A) did not show it. This may be due to a DFT error on the height of the fluorine in relation to the silver neighbors. Another possibility is that DFT overestimates the decaying length of the $2p$ orbital. For example, self-interaction errors~\cite{Filippetti2009} tend to yield larger decay lengths. In contrast, if the orbital were more localized the protrusion would be buried in the Ag depression. Still another possibility is that the F does not occupy a high symmetry position, as assumed, but is displaced laterally from the short bridge position lowering the height.

The simulations underestimated the lateral size of the wells and overestimated the depth. This issue was attributed to the various simplifications of the Tersoff-Hamann approximation.  Furthermore, lateral resolution effects, inherent to the tip shape can play an important role in these problems.
We hope the present results will stimulate more theoretical and experimental work to shed light on the AT(C) assignment and the bias dependence of AT(A). 

We have introduced a simplified orbital model that provides an intuitive understanding of the ATs and allows us to link them to important parameters of the adsorption and the reaction occurring on the surface. Using a simple estimate for the orbital extension based on the classical work by Slater we argued that the evanescent wave of F $2p$ and Ag $5s$ orbitals have similar strength at the typical tip distance. Thus, two different features associated with these orbitals, namely the protrusion (F $2p$) and/or the depression (Ag $5s$) can appear depending on the parameters.  The depression is attributed to the oxidation of the nearby Ag atoms by F. This produces a decrease of the screening charge in Ag and a contraction of the $5s$ orbitals which appears as an apparent topographic depression.  The protrusion is attributed to the filled $2p$ orbitals of the \ch{F^-} ion. This explains why the protrusion tends to appear at the occupied part of the spectrum (negative bias).

From a broader perspective, the present orbital model provides a clue to understanding the ATs, offering a complementary approach to other methods that do not focus on orbitals~\cite{Lang1986,Lang1987,Sautet1997}. This approach may prove useful not only in the present context but also in any system where detailed surface chemistry information is needed~\cite{Roman2014,Besenbacher1993,Andryushechkin2018b,Grochala2001,McLain2006,Grochala2006a,Yang2014,Gawraczynski2019,Grzelak2020,Miller2020,Sanchez-Movellan2021,Bachar2022,Piombo2022,Prosnikov2022,Wilkinson2023,Wintterlin2000,Jones1988,Altman2001, Serafin1998}. Indeed, the model suggests that STM can be used to evaluate the local valence of metallic ions which can have a wide range of applications. For example, STM studies of catalytic reactions~\cite{Wintterlin2000,Jones1988,Altman2001, Serafin1998} can greatly benefit from local chemical information. More theoretical and experimental work is needed to explore this idea in different situations. In particular, this calls for systematic STM studies combined with techniques that give access to chemical information in transition metals exposed to halogens and chalcogens.

	\section{\label{sec:ACKNOWLEDGMENTS}ACKNOWLEDGMENTS}
We are in debt to Yanina Fasano, Armando Aligia, and Rosanna Larciprete for enlightening discussions. Work supported by the Italian Ministry of University and Research through the project Quantum Transition-metal FLUOrides (QT-FLUO) PRIN 20207ZXT4Z and CNR-CONICET NMES project. We acknowledge the CINECA award under the ISCRA initiative, for the availability of high-performance computing resources and support through class C projects ABRID (HP10CKR6YT) and Frage (HP10CDJZ0O) and class B project FluSCA (HP10BAWWVS). Polish authors appreciate support from the National Science Center (NCN), project Maestro (2017/26/A/ST5/00570).
\appendix
\section{Computation of the depression depth \label{sec:app}}

Here we present a computation of the apparent depression depth in terms of the simplified model. As in the main text, we use 0 and $\infty$ to label quantities at the F and pristine sites, respectively (e.g. $I_0(z)\equiv I(z)|_{x=y=0}$, etc.).

We can restrict the sum in Eq.~\eqref{eq:TH} to the $\zeta$ planar nearest neighbor (nn) of the F adatom. They contribute the same density and have the same cumulative DOS at $x=y=0$ namely, $\rho_{05s}(z)$ and $G_{05s}$, and analogously for the pristine site. With these definitions, the current at both sites reads, 
\begin{eqnarray}\label{eq:TH0}
    I_0(z_0,V_{\rm B})&\approx&\frac{4\pi e}\hbar|M|^{2}N_t(0) \zeta  G_{05s}(V_{\rm B}) \rho_{0  5s}(z_0),\\
    I_\infty(z_\infty,V_{\rm B})&\approx&\frac{4\pi e}\hbar|M|^{2}N_t(0) \zeta  G_{\infty5s}(V_{\rm B}) \rho_{\infty  5s}(z_\infty).
\end{eqnarray}
We next expand the current in the fluorine site to linear order in the wall height, i.e.   $I_0(z_0,V_{\rm B})=I_0(z_\infty,V_{\rm B})+(\partial I_0/\partial z)\Delta z$ and define the constant height differences $\Delta I(z_\infty,V_{\rm B})=I_0(z_\infty,V_{\rm B})-I_\infty(z_\infty,V_{\rm B})$, etc. Keeping only linear terms in the variations one obtains, 
\begin{eqnarray}\label{eq:deltaI}
    \Delta && I(z_\infty,V_{\rm B})\approx\frac{4\pi e}\hbar|M|^{2}N_t(0)\zeta\\
    && \times\left[ \Delta G_{5s}(V_{\rm B}) \rho_{{\infty}  5s}(z_\infty)+ G_{{\infty}5s}(V_{\rm B}) \Delta \rho_{5s}(z_\infty)\right].\nonumber
\end{eqnarray}

Imposing the condition of constant current,  $I_0(z_0,V_{\rm B})=I_\infty(z_\infty,V_{\rm B})$, defines the variation of height $\Delta z$ due to the presence of the F and leads to Eq.~\eqref{eq:dzdr}.

To evaluate the equation we need the distance dependence of the charge. We evaluate this quantity using hydrogenic wave functions and assuming that the main factor determining the change in $\rho_{{0} 5s}$ is the change of the Ag oxidation state due to the charge transfer to the F atom. Other effects, such as the deformation of the atomic cloud due to the presence of the \ch{F^-} ion, are neglected. We provide the expression of the charge for general hydrogenic orbitals. For large distances, the hydrogenic charge distribution can be written as~\cite{Slater1930}, 
\begin{eqnarray}\label{eq:rhonl}
    && \rho_{nl}(r)=2(2l+1)\\
    && \times \left(\frac{2r Z_{\rm eff}^{nl}}{n a_0}\right)^{2 n-2}\left(\frac{Z_{\rm eff}^{nl}}{ a_0}\right)^3 \frac{ \exp({-{2 r Z_{\rm eff}^{nl}}/{n a_0 }})}{\pi n^4 (n-l-1)! (l+n)!}.\nonumber
\end{eqnarray}
Performing the derivative and inserting the expressions in Eq.~\eqref{eq:dzdr} leads to Eq.~\eqref{eq:dzdz}.


\begin{thebibliography}{72}%
\makeatletter
\providecommand \@ifxundefined [1]{%
 \@ifx{#1\undefined}
}%
\providecommand \@ifnum [1]{%
 \ifnum #1\expandafter \@firstoftwo
 \else \expandafter \@secondoftwo
 \fi
}%
\providecommand \@ifx [1]{%
 \ifx #1\expandafter \@firstoftwo
 \else \expandafter \@secondoftwo
 \fi
}%
\providecommand \natexlab [1]{#1}%
\providecommand \enquote  [1]{``#1''}%
\providecommand \bibnamefont  [1]{#1}%
\providecommand \bibfnamefont [1]{#1}%
\providecommand \citenamefont [1]{#1}%
\providecommand \href@noop [0]{\@secondoftwo}%
\providecommand \href [0]{\begingroup \@sanitize@url \@href}%
\providecommand \@href[1]{\@@startlink{#1}\@@href}%
\providecommand \@@href[1]{\endgroup#1\@@endlink}%
\providecommand \@sanitize@url [0]{\catcode `\\12\catcode `\$12\catcode
  `\&12\catcode `\#12\catcode `\^12\catcode `\_12\catcode `\%12\relax}%
\providecommand \@@startlink[1]{}%
\providecommand \@@endlink[0]{}%
\providecommand \url  [0]{\begingroup\@sanitize@url \@url }%
\providecommand \@url [1]{\endgroup\@href {#1}{\urlprefix }}%
\providecommand \urlprefix  [0]{URL }%
\providecommand \Eprint [0]{\href }%
\providecommand \doibase [0]{https://doi.org/}%
\providecommand \selectlanguage [0]{\@gobble}%
\providecommand \bibinfo  [0]{\@secondoftwo}%
\providecommand \bibfield  [0]{\@secondoftwo}%
\providecommand \translation [1]{[#1]}%
\providecommand \BibitemOpen [0]{}%
\providecommand \bibitemStop [0]{}%
\providecommand \bibitemNoStop [0]{.\EOS\space}%
\providecommand \EOS [0]{\spacefactor3000\relax}%
\providecommand \BibitemShut  [1]{\csname bibitem#1\endcsname}%
\let\auto@bib@innerbib\@empty
\bibitem [{\citenamefont {Roman}\ \emph {et~al.}(2014)\citenamefont {Roman},
  \citenamefont {Gossenberger}, \citenamefont {Forster-Tonigold},\ and\
  \citenamefont {Gro{\ss}}}]{Roman2014}%
  \BibitemOpen
  \bibfield  {author} {\bibinfo {author} {\bibfnamefont {T.}~\bibnamefont
  {Roman}}, \bibinfo {author} {\bibfnamefont {F.}~\bibnamefont {Gossenberger}},
  \bibinfo {author} {\bibfnamefont {K.}~\bibnamefont {Forster-Tonigold}},\ and\
  \bibinfo {author} {\bibfnamefont {A.}~\bibnamefont {Gro{\ss}}},\ }\href
  {https://doi.org/10.1039/C4CP00237G} {\bibfield  {journal} {\bibinfo
  {journal} {Phys. Chem. Chem. Phys.}\ }\textbf {\bibinfo {volume} {16}},\
  \bibinfo {pages} {13630} (\bibinfo {year} {2014})}\BibitemShut {NoStop}%
\bibitem [{\citenamefont {Besenbacher}\ and\ \citenamefont
  {N{\o}rskov}(1993)}]{Besenbacher1993}%
  \BibitemOpen
  \bibfield  {author} {\bibinfo {author} {\bibfnamefont {F.}~\bibnamefont
  {Besenbacher}}\ and\ \bibinfo {author} {\bibfnamefont {J.~K.}\ \bibnamefont
  {N{\o}rskov}},\ }\href {https://doi.org/10.1016/0079-6816(93)90006-H}
  {\bibfield  {journal} {\bibinfo  {journal} {Prog. Surf. Sci.}\ }\textbf
  {\bibinfo {volume} {44}},\ \bibinfo {pages} {5} (\bibinfo {year}
  {1993})}\BibitemShut {NoStop}%
\bibitem [{\citenamefont {Andryushechkin}\ \emph {et~al.}(2018)\citenamefont
  {Andryushechkin}, \citenamefont {Pavlova},\ and\ \citenamefont
  {Eltsov}}]{Andryushechkin2018b}%
  \BibitemOpen
  \bibfield  {author} {\bibinfo {author} {\bibfnamefont {B.~V.}\ \bibnamefont
  {Andryushechkin}}, \bibinfo {author} {\bibfnamefont {T.~V.}\ \bibnamefont
  {Pavlova}},\ and\ \bibinfo {author} {\bibfnamefont {K.~N.}\ \bibnamefont
  {Eltsov}},\ }\href {https://doi.org/10.1016/j.surfrep.2018.03.001} {\bibfield
   {journal} {\bibinfo  {journal} {Surf. Sci. Rep.}\ }\textbf {\bibinfo
  {volume} {73}},\ \bibinfo {pages} {83} (\bibinfo {year} {2018})}\BibitemShut
  {NoStop}%
\bibitem [{\citenamefont {Lin}\ \emph {et~al.}(2021)\citenamefont {Lin},
  \citenamefont {{Villar Arribi}}, \citenamefont {Fabbris}, \citenamefont
  {Botana}, \citenamefont {Meyers}, \citenamefont {Miao}, \citenamefont {Shen},
  \citenamefont {Mazzone}, \citenamefont {Feng}, \citenamefont {Chiuzbǎian},
  \citenamefont {Nag}, \citenamefont {Walters}, \citenamefont
  {Garc{\'{i}}a-Fern{\'{a}}ndez}, \citenamefont {Zhou}, \citenamefont
  {Pelliciari}, \citenamefont {Jarrige}, \citenamefont {Freeland},
  \citenamefont {Zhang}, \citenamefont {Mitchell}, \citenamefont {Bisogni},
  \citenamefont {Liu}, \citenamefont {Norman},\ and\ \citenamefont
  {Dean}}]{Lin2021}%
  \BibitemOpen
  \bibfield  {author} {\bibinfo {author} {\bibfnamefont {J.~Q.}\ \bibnamefont
  {Lin}}, \bibinfo {author} {\bibfnamefont {P.}~\bibnamefont {{Villar
  Arribi}}}, \bibinfo {author} {\bibfnamefont {G.}~\bibnamefont {Fabbris}},
  \bibinfo {author} {\bibfnamefont {A.~S.}\ \bibnamefont {Botana}}, \bibinfo
  {author} {\bibfnamefont {D.}~\bibnamefont {Meyers}}, \bibinfo {author}
  {\bibfnamefont {H.}~\bibnamefont {Miao}}, \bibinfo {author} {\bibfnamefont
  {Y.}~\bibnamefont {Shen}}, \bibinfo {author} {\bibfnamefont {D.~G.}\
  \bibnamefont {Mazzone}}, \bibinfo {author} {\bibfnamefont {J.}~\bibnamefont
  {Feng}}, \bibinfo {author} {\bibfnamefont {S.~G.}\ \bibnamefont
  {Chiuzbǎian}}, \bibinfo {author} {\bibfnamefont {A.}~\bibnamefont {Nag}},
  \bibinfo {author} {\bibfnamefont {A.~C.}\ \bibnamefont {Walters}}, \bibinfo
  {author} {\bibfnamefont {M.}~\bibnamefont {Garc{\'{i}}a-Fern{\'{a}}ndez}},
  \bibinfo {author} {\bibfnamefont {K.~J.}\ \bibnamefont {Zhou}}, \bibinfo
  {author} {\bibfnamefont {J.}~\bibnamefont {Pelliciari}}, \bibinfo {author}
  {\bibfnamefont {I.}~\bibnamefont {Jarrige}}, \bibinfo {author} {\bibfnamefont
  {J.~W.}\ \bibnamefont {Freeland}}, \bibinfo {author} {\bibfnamefont
  {J.}~\bibnamefont {Zhang}}, \bibinfo {author} {\bibfnamefont {J.~F.}\
  \bibnamefont {Mitchell}}, \bibinfo {author} {\bibfnamefont {V.}~\bibnamefont
  {Bisogni}}, \bibinfo {author} {\bibfnamefont {X.}~\bibnamefont {Liu}},
  \bibinfo {author} {\bibfnamefont {M.~R.}\ \bibnamefont {Norman}},\ and\
  \bibinfo {author} {\bibfnamefont {M.~P.}\ \bibnamefont {Dean}},\ }\href
  {https://doi.org/10.1103/PhysRevLett.126.087001} {\bibfield  {journal}
  {\bibinfo  {journal} {Phys. Rev. Lett.}\ }\textbf {\bibinfo {volume} {126}},\
  \bibinfo {pages} {087001} (\bibinfo {year} {2021})},\ \Eprint
  {https://arxiv.org/abs/2008.08209} {arXiv:2008.08209} \BibitemShut {NoStop}%
\bibitem [{\citenamefont {Grochala}\ and\ \citenamefont
  {Hoffmann}(2001)}]{Grochala2001}%
  \BibitemOpen
  \bibfield  {author} {\bibinfo {author} {\bibfnamefont {W.}~\bibnamefont
  {Grochala}}\ and\ \bibinfo {author} {\bibfnamefont {R.}~\bibnamefont
  {Hoffmann}},\ }\href
  {https://doi.org/10.1002/1521-3773(20010803)40:15<2742::AID-ANIE2742>3.0.CO;2-X}
  {\bibfield  {journal} {\bibinfo  {journal} {Angew. Chemie Int. Ed.}\ }\textbf
  {\bibinfo {volume} {40}},\ \bibinfo {pages} {2742} (\bibinfo {year}
  {2001})}\BibitemShut {NoStop}%
\bibitem [{\citenamefont {McLain}\ \emph {et~al.}(2006)\citenamefont {McLain},
  \citenamefont {Dolgos}, \citenamefont {Tennant}, \citenamefont {Turner},
  \citenamefont {Barnes}, \citenamefont {Proffen}, \citenamefont {Sales},\ and\
  \citenamefont {Bewley}}]{McLain2006}%
  \BibitemOpen
  \bibfield  {author} {\bibinfo {author} {\bibfnamefont {S.~E.}\ \bibnamefont
  {McLain}}, \bibinfo {author} {\bibfnamefont {M.~R.}\ \bibnamefont {Dolgos}},
  \bibinfo {author} {\bibfnamefont {D.~A.}\ \bibnamefont {Tennant}}, \bibinfo
  {author} {\bibfnamefont {J.~F.~C.}\ \bibnamefont {Turner}}, \bibinfo {author}
  {\bibfnamefont {T.}~\bibnamefont {Barnes}}, \bibinfo {author} {\bibfnamefont
  {T.}~\bibnamefont {Proffen}}, \bibinfo {author} {\bibfnamefont {B.~C.}\
  \bibnamefont {Sales}},\ and\ \bibinfo {author} {\bibfnamefont {R.~I.}\
  \bibnamefont {Bewley}},\ }\href {https://doi.org/10.1038/nmat1670} {\bibfield
   {journal} {\bibinfo  {journal} {Nat. Mater.}\ }\textbf {\bibinfo {volume}
  {5}},\ \bibinfo {pages} {561} (\bibinfo {year} {2006})}\BibitemShut {NoStop}%
\bibitem [{\citenamefont {Grochala}(2006)}]{Grochala2006a}%
  \BibitemOpen
  \bibfield  {author} {\bibinfo {author} {\bibfnamefont {W.}~\bibnamefont
  {Grochala}},\ }\href {https://doi.org/10.1038/nmat1678} {\bibfield  {journal}
  {\bibinfo  {journal} {Nat. Mater.}\ }\textbf {\bibinfo {volume} {5}},\
  \bibinfo {pages} {513} (\bibinfo {year} {2006})}\BibitemShut {NoStop}%
\bibitem [{\citenamefont {Yang}\ and\ \citenamefont {Su}(2015)}]{Yang2014}%
  \BibitemOpen
  \bibfield  {author} {\bibinfo {author} {\bibfnamefont {X.}~\bibnamefont
  {Yang}}\ and\ \bibinfo {author} {\bibfnamefont {H.}~\bibnamefont {Su}},\
  }\href {https://doi.org/10.1038/srep05420} {\bibfield  {journal} {\bibinfo
  {journal} {Sci. Rep.}\ }\textbf {\bibinfo {volume} {4}},\ \bibinfo {pages}
  {5420} (\bibinfo {year} {2015})}\BibitemShut {NoStop}%
\bibitem [{\citenamefont {Gawraczy{\'{n}}ski}\ \emph
  {et~al.}(2019)\citenamefont {Gawraczy{\'{n}}ski}, \citenamefont
  {Kurzyd{\l}owski}, \citenamefont {Ewings}, \citenamefont {Bandaru},
  \citenamefont {Gadomski}, \citenamefont {Mazej}, \citenamefont {Ruani},
  \citenamefont {Bergenti}, \citenamefont {Jaro{\'{n}}}, \citenamefont
  {Ozarowski}, \citenamefont {Hill}, \citenamefont {Leszczy{\'{n}}ski},
  \citenamefont {Tok{\'{a}}r}, \citenamefont {Derzsi}, \citenamefont {Barone},
  \citenamefont {Wohlfeld}, \citenamefont {Lorenzana},\ and\ \citenamefont
  {Grochala}}]{Gawraczynski2019}%
  \BibitemOpen
  \bibfield  {author} {\bibinfo {author} {\bibfnamefont {J.}~\bibnamefont
  {Gawraczy{\'{n}}ski}}, \bibinfo {author} {\bibfnamefont {D.}~\bibnamefont
  {Kurzyd{\l}owski}}, \bibinfo {author} {\bibfnamefont {R.~A.}\ \bibnamefont
  {Ewings}}, \bibinfo {author} {\bibfnamefont {S.}~\bibnamefont {Bandaru}},
  \bibinfo {author} {\bibfnamefont {W.}~\bibnamefont {Gadomski}}, \bibinfo
  {author} {\bibfnamefont {Z.}~\bibnamefont {Mazej}}, \bibinfo {author}
  {\bibfnamefont {G.}~\bibnamefont {Ruani}}, \bibinfo {author} {\bibfnamefont
  {I.}~\bibnamefont {Bergenti}}, \bibinfo {author} {\bibfnamefont
  {T.}~\bibnamefont {Jaro{\'{n}}}}, \bibinfo {author} {\bibfnamefont
  {A.}~\bibnamefont {Ozarowski}}, \bibinfo {author} {\bibfnamefont
  {S.}~\bibnamefont {Hill}}, \bibinfo {author} {\bibfnamefont {P.~J.}\
  \bibnamefont {Leszczy{\'{n}}ski}}, \bibinfo {author} {\bibfnamefont
  {K.}~\bibnamefont {Tok{\'{a}}r}}, \bibinfo {author} {\bibfnamefont
  {M.}~\bibnamefont {Derzsi}}, \bibinfo {author} {\bibfnamefont
  {P.}~\bibnamefont {Barone}}, \bibinfo {author} {\bibfnamefont
  {K.}~\bibnamefont {Wohlfeld}}, \bibinfo {author} {\bibfnamefont
  {J.}~\bibnamefont {Lorenzana}},\ and\ \bibinfo {author} {\bibfnamefont
  {W.}~\bibnamefont {Grochala}},\ }\href
  {https://doi.org/10.1073/pnas.1812857116} {\bibfield  {journal} {\bibinfo
  {journal} {Proc. Natl. Acad. Sci. U. S. A.}\ }\textbf {\bibinfo {volume}
  {116}},\ \bibinfo {pages} {1495} (\bibinfo {year} {2019})},\ \Eprint
  {https://arxiv.org/abs/1804.00329} {arXiv:1804.00329} \BibitemShut {NoStop}%
\bibitem [{\citenamefont {Grzelak}\ \emph {et~al.}(2020)\citenamefont
  {Grzelak}, \citenamefont {Su}, \citenamefont {Yang}, \citenamefont
  {Kurzyd{\l}owski}, \citenamefont {Lorenzana},\ and\ \citenamefont
  {Grochala}}]{Grzelak2020}%
  \BibitemOpen
  \bibfield  {author} {\bibinfo {author} {\bibfnamefont {A.}~\bibnamefont
  {Grzelak}}, \bibinfo {author} {\bibfnamefont {H.}~\bibnamefont {Su}},
  \bibinfo {author} {\bibfnamefont {X.}~\bibnamefont {Yang}}, \bibinfo {author}
  {\bibfnamefont {D.}~\bibnamefont {Kurzyd{\l}owski}}, \bibinfo {author}
  {\bibfnamefont {J.}~\bibnamefont {Lorenzana}},\ and\ \bibinfo {author}
  {\bibfnamefont {W.}~\bibnamefont {Grochala}},\ }\href
  {https://doi.org/10.1103/physrevmaterials.4.084405} {\bibfield  {journal}
  {\bibinfo  {journal} {Phys. Rev. Mater.}\ }\textbf {\bibinfo {volume} {4}},\
  \bibinfo {pages} {084405} (\bibinfo {year} {2020})},\ \Eprint
  {https://arxiv.org/abs/2005.00461} {arXiv:2005.00461} \BibitemShut {NoStop}%
\bibitem [{\citenamefont {Miller}\ and\ \citenamefont
  {Botana}(2020)}]{Miller2020}%
  \BibitemOpen
  \bibfield  {author} {\bibinfo {author} {\bibfnamefont {C.}~\bibnamefont
  {Miller}}\ and\ \bibinfo {author} {\bibfnamefont {A.~S.}\ \bibnamefont
  {Botana}},\ }\href {https://doi.org/10.1103/PhysRevB.101.195116} {\bibfield
  {journal} {\bibinfo  {journal} {Phys. Rev. B}\ }\textbf {\bibinfo {volume}
  {101}},\ \bibinfo {pages} {195116} (\bibinfo {year} {2020})}\BibitemShut
  {NoStop}%
\bibitem [{\citenamefont {S{\'{a}}nchez-Movell{\'{a}}n}\ \emph
  {et~al.}(2021)\citenamefont {S{\'{a}}nchez-Movell{\'{a}}n}, \citenamefont
  {Moreno-Ceballos}, \citenamefont {Garc{\'{i}}a-Fern{\'{a}}ndez},
  \citenamefont {Aramburu},\ and\ \citenamefont
  {Moreno}}]{Sanchez-Movellan2021}%
  \BibitemOpen
  \bibfield  {author} {\bibinfo {author} {\bibfnamefont {I.}~\bibnamefont
  {S{\'{a}}nchez-Movell{\'{a}}n}}, \bibinfo {author} {\bibfnamefont
  {J.}~\bibnamefont {Moreno-Ceballos}}, \bibinfo {author} {\bibfnamefont
  {P.}~\bibnamefont {Garc{\'{i}}a-Fern{\'{a}}ndez}}, \bibinfo {author}
  {\bibfnamefont {J.~A.}\ \bibnamefont {Aramburu}},\ and\ \bibinfo {author}
  {\bibfnamefont {M.}~\bibnamefont {Moreno}},\ }\href
  {https://doi.org/10.1002/chem.202101865} {\bibfield  {journal} {\bibinfo
  {journal} {Chem. – A Eur. J.}\ ,\ \bibinfo {pages} {1}} (\bibinfo {year}
  {2021})}\BibitemShut {NoStop}%
\bibitem [{\citenamefont {Bachar}\ \emph {et~al.}(2022)\citenamefont {Bachar},
  \citenamefont {Koteras}, \citenamefont {Gawraczynski}, \citenamefont
  {Trzci{\'{n}}ski}, \citenamefont {Paszula}, \citenamefont {Piombo},
  \citenamefont {Barone}, \citenamefont {Mazej}, \citenamefont {Ghiringhelli},
  \citenamefont {Nag}, \citenamefont {Zhou}, \citenamefont {Lorenzana},
  \citenamefont {van~der Marel},\ and\ \citenamefont {Grochala}}]{Bachar2022}%
  \BibitemOpen
  \bibfield  {author} {\bibinfo {author} {\bibfnamefont {N.}~\bibnamefont
  {Bachar}}, \bibinfo {author} {\bibfnamefont {K.}~\bibnamefont {Koteras}},
  \bibinfo {author} {\bibfnamefont {J.}~\bibnamefont {Gawraczynski}}, \bibinfo
  {author} {\bibfnamefont {W.}~\bibnamefont {Trzci{\'{n}}ski}}, \bibinfo
  {author} {\bibfnamefont {J.}~\bibnamefont {Paszula}}, \bibinfo {author}
  {\bibfnamefont {R.}~\bibnamefont {Piombo}}, \bibinfo {author} {\bibfnamefont
  {P.}~\bibnamefont {Barone}}, \bibinfo {author} {\bibfnamefont
  {Z.}~\bibnamefont {Mazej}}, \bibinfo {author} {\bibfnamefont
  {G.}~\bibnamefont {Ghiringhelli}}, \bibinfo {author} {\bibfnamefont
  {A.}~\bibnamefont {Nag}}, \bibinfo {author} {\bibfnamefont {K.-j.}\
  \bibnamefont {Zhou}}, \bibinfo {author} {\bibfnamefont {J.}~\bibnamefont
  {Lorenzana}}, \bibinfo {author} {\bibfnamefont {D.}~\bibnamefont {van~der
  Marel}},\ and\ \bibinfo {author} {\bibfnamefont {W.}~\bibnamefont
  {Grochala}},\ }\href {https://doi.org/10.1103/PhysRevResearch.4.023108}
  {\bibfield  {journal} {\bibinfo  {journal} {Phys. Rev. Res.}\ }\textbf
  {\bibinfo {volume} {4}},\ \bibinfo {pages} {023108} (\bibinfo {year}
  {2022})},\ \Eprint {https://arxiv.org/abs/2105.08862} {arXiv:2105.08862}
  \BibitemShut {NoStop}%
\bibitem [{\citenamefont {Piombo}\ \emph {et~al.}(2022)\citenamefont {Piombo},
  \citenamefont {Jezierski}, \citenamefont {Martins}, \citenamefont
  {Jaro{\'{n}}}, \citenamefont {Gastiasoro}, \citenamefont {Barone},
  \citenamefont {Tok{\'{a}}r}, \citenamefont {Piekarz}, \citenamefont {Derzsi},
  \citenamefont {Mazej}, \citenamefont {Abbate}, \citenamefont {Grochala},\
  and\ \citenamefont {Lorenzana}}]{Piombo2022}%
  \BibitemOpen
  \bibfield  {author} {\bibinfo {author} {\bibfnamefont {R.}~\bibnamefont
  {Piombo}}, \bibinfo {author} {\bibfnamefont {D.}~\bibnamefont {Jezierski}},
  \bibinfo {author} {\bibfnamefont {H.~P.}\ \bibnamefont {Martins}}, \bibinfo
  {author} {\bibfnamefont {T.}~\bibnamefont {Jaro{\'{n}}}}, \bibinfo {author}
  {\bibfnamefont {M.~N.}\ \bibnamefont {Gastiasoro}}, \bibinfo {author}
  {\bibfnamefont {P.}~\bibnamefont {Barone}}, \bibinfo {author} {\bibfnamefont
  {K.}~\bibnamefont {Tok{\'{a}}r}}, \bibinfo {author} {\bibfnamefont
  {P.}~\bibnamefont {Piekarz}}, \bibinfo {author} {\bibfnamefont
  {M.}~\bibnamefont {Derzsi}}, \bibinfo {author} {\bibfnamefont
  {Z.}~\bibnamefont {Mazej}}, \bibinfo {author} {\bibfnamefont
  {M.}~\bibnamefont {Abbate}}, \bibinfo {author} {\bibfnamefont
  {W.}~\bibnamefont {Grochala}},\ and\ \bibinfo {author} {\bibfnamefont
  {J.}~\bibnamefont {Lorenzana}},\ }\href
  {https://doi.org/10.1103/PhysRevB.106.035142} {\bibfield  {journal} {\bibinfo
   {journal} {Phys. Rev. B}\ }\textbf {\bibinfo {volume} {106}},\ \bibinfo
  {pages} {035142} (\bibinfo {year} {2022})}\BibitemShut {NoStop}%
\bibitem [{\citenamefont {Prosnikov}(2022)}]{Prosnikov2022}%
  \BibitemOpen
  \bibfield  {author} {\bibinfo {author} {\bibfnamefont {M.~A.}\ \bibnamefont
  {Prosnikov}},\ }\bibfield  {journal} {\bibinfo  {journal} {J. Magn. Magn.
  Mater.}\ }\textbf {\bibinfo {volume} {557}},\ \href
  {https://doi.org/10.1016/j.jmmm.2022.169432} {10.1016/j.jmmm.2022.169432}
  (\bibinfo {year} {2022})\BibitemShut {NoStop}%
\bibitem [{\citenamefont {Wilkinson}\ \emph {et~al.}(2023)\citenamefont
  {Wilkinson}, \citenamefont {Blundell}, \citenamefont {Biesenkamp},
  \citenamefont {Braden}, \citenamefont {Hansen}, \citenamefont {Koteras},
  \citenamefont {Grochala}, \citenamefont {Barone}, \citenamefont {Lorenzana},
  \citenamefont {Mazej},\ and\ \citenamefont {Tav{\v{c}}ar}}]{Wilkinson2023}%
  \BibitemOpen
  \bibfield  {author} {\bibinfo {author} {\bibfnamefont {J.~M.}\ \bibnamefont
  {Wilkinson}}, \bibinfo {author} {\bibfnamefont {S.~J.}\ \bibnamefont
  {Blundell}}, \bibinfo {author} {\bibfnamefont {S.}~\bibnamefont
  {Biesenkamp}}, \bibinfo {author} {\bibfnamefont {M.}~\bibnamefont {Braden}},
  \bibinfo {author} {\bibfnamefont {T.~C.}\ \bibnamefont {Hansen}}, \bibinfo
  {author} {\bibfnamefont {K.}~\bibnamefont {Koteras}}, \bibinfo {author}
  {\bibfnamefont {W.}~\bibnamefont {Grochala}}, \bibinfo {author}
  {\bibfnamefont {P.}~\bibnamefont {Barone}}, \bibinfo {author} {\bibfnamefont
  {J.}~\bibnamefont {Lorenzana}}, \bibinfo {author} {\bibfnamefont
  {Z.}~\bibnamefont {Mazej}},\ and\ \bibinfo {author} {\bibfnamefont
  {G.}~\bibnamefont {Tav{\v{c}}ar}},\ }\href
  {https://doi.org/10.1103/PhysRevB.107.144422} {\bibfield  {journal} {\bibinfo
   {journal} {Phys. Rev. B}\ }\textbf {\bibinfo {volume} {107}},\ \bibinfo
  {pages} {144422} (\bibinfo {year} {2023})}\BibitemShut {NoStop}%
\bibitem [{\citenamefont {Ignaczak}\ and\ \citenamefont
  {Gomes}(1997)}]{Ignaczak1997}%
  \BibitemOpen
  \bibfield  {author} {\bibinfo {author} {\bibfnamefont {A.}~\bibnamefont
  {Ignaczak}}\ and\ \bibinfo {author} {\bibfnamefont {J.~A.}\ \bibnamefont
  {Gomes}},\ }\href {https://doi.org/10.1016/S0022-0728(96)04815-2} {\bibfield
  {journal} {\bibinfo  {journal} {J. Electroanal. Chem.}\ }\textbf {\bibinfo
  {volume} {420}},\ \bibinfo {pages} {71} (\bibinfo {year} {1997})}\BibitemShut
  {NoStop}%
\bibitem [{\citenamefont {Tripkovic}\ \emph {et~al.}(2009)\citenamefont
  {Tripkovic}, \citenamefont {Strmcnik}, \citenamefont {van~der Vliet},
  \citenamefont {Stamenkovic},\ and\ \citenamefont {Markovic}}]{Tripkovic2009}%
  \BibitemOpen
  \bibfield  {author} {\bibinfo {author} {\bibfnamefont {D.~V.}\ \bibnamefont
  {Tripkovic}}, \bibinfo {author} {\bibfnamefont {D.}~\bibnamefont {Strmcnik}},
  \bibinfo {author} {\bibfnamefont {D.}~\bibnamefont {van~der Vliet}}, \bibinfo
  {author} {\bibfnamefont {V.}~\bibnamefont {Stamenkovic}},\ and\ \bibinfo
  {author} {\bibfnamefont {N.~M.}\ \bibnamefont {Markovic}},\ }\href
  {https://doi.org/10.1039/B803714K} {\bibfield  {journal} {\bibinfo  {journal}
  {Faraday Discuss.}\ }\textbf {\bibinfo {volume} {140}},\ \bibinfo {pages}
  {25} (\bibinfo {year} {2009})}\BibitemShut {NoStop}%
\bibitem [{\citenamefont {Zhu}\ and\ \citenamefont {Wang}(2016)}]{Zhu2016}%
  \BibitemOpen
  \bibfield  {author} {\bibinfo {author} {\bibfnamefont {Q.}~\bibnamefont
  {Zhu}}\ and\ \bibinfo {author} {\bibfnamefont {S.-q.}\ \bibnamefont {Wang}},\
  }\href {https://doi.org/10.1149/2.0821609jes} {\bibfield  {journal} {\bibinfo
   {journal} {J. Electrochem. Soc.}\ }\textbf {\bibinfo {volume} {163}},\
  \bibinfo {pages} {H796} (\bibinfo {year} {2016})}\BibitemShut {NoStop}%
\bibitem [{\citenamefont {Zaum}\ and\ \citenamefont
  {Morgenstern}(2018)}]{Zaum2018}%
  \BibitemOpen
  \bibfield  {author} {\bibinfo {author} {\bibfnamefont {C.}~\bibnamefont
  {Zaum}}\ and\ \bibinfo {author} {\bibfnamefont {K.}~\bibnamefont
  {Morgenstern}},\ }\href {https://doi.org/10.1063/1.5032174} {\bibfield
  {journal} {\bibinfo  {journal} {Appl. Phys. Lett.}\ }\textbf {\bibinfo
  {volume} {113}},\ \bibinfo {pages} {31602} (\bibinfo {year}
  {2018})}\BibitemShut {NoStop}%
\bibitem [{\citenamefont {Spurgeon}\ \emph {et~al.}(2019)\citenamefont
  {Spurgeon}, \citenamefont {Liu}, \citenamefont {Walen}, \citenamefont {Oh},
  \citenamefont {Yang}, \citenamefont {Kim},\ and\ \citenamefont
  {Thiel}}]{Spurgeon2019}%
  \BibitemOpen
  \bibfield  {author} {\bibinfo {author} {\bibfnamefont {P.~M.}\ \bibnamefont
  {Spurgeon}}, \bibinfo {author} {\bibfnamefont {D.~J.}\ \bibnamefont {Liu}},
  \bibinfo {author} {\bibfnamefont {H.}~\bibnamefont {Walen}}, \bibinfo
  {author} {\bibfnamefont {J.}~\bibnamefont {Oh}}, \bibinfo {author}
  {\bibfnamefont {H.~J.}\ \bibnamefont {Yang}}, \bibinfo {author}
  {\bibfnamefont {Y.}~\bibnamefont {Kim}},\ and\ \bibinfo {author}
  {\bibfnamefont {P.~A.}\ \bibnamefont {Thiel}},\ }\href
  {https://doi.org/10.1039/c9cp01626k} {\bibfield  {journal} {\bibinfo
  {journal} {Phys. Chem. Chem. Phys.}\ }\textbf {\bibinfo {volume} {21}},\
  \bibinfo {pages} {10540} (\bibinfo {year} {2019})}\BibitemShut {NoStop}%
\bibitem [{\citenamefont {Lee}\ \emph {et~al.}(2020)\citenamefont {Lee},
  \citenamefont {Lee}, \citenamefont {Palot{\'{a}}s}, \citenamefont {Lee},\
  and\ \citenamefont {Soon}}]{Lee2020}%
  \BibitemOpen
  \bibfield  {author} {\bibinfo {author} {\bibfnamefont {G.}~\bibnamefont
  {Lee}}, \bibinfo {author} {\bibfnamefont {Y.-J.}\ \bibnamefont {Lee}},
  \bibinfo {author} {\bibfnamefont {K.}~\bibnamefont {Palot{\'{a}}s}}, \bibinfo
  {author} {\bibfnamefont {T.}~\bibnamefont {Lee}},\ and\ \bibinfo {author}
  {\bibfnamefont {A.}~\bibnamefont {Soon}},\ }\href
  {https://doi.org/10.1021/acs.jpcc.0c02842} {\bibfield  {journal} {\bibinfo
  {journal} {J. Phys. Chem. C}\ }\textbf {\bibinfo {volume} {124}},\ \bibinfo
  {pages} {16362} (\bibinfo {year} {2020})}\BibitemShut {NoStop}%
\bibitem [{\citenamefont {Li}\ \emph {et~al.}(2002)\citenamefont {Li},
  \citenamefont {Stampfl},\ and\ \citenamefont {Scheffler}}]{Li2002a}%
  \BibitemOpen
  \bibfield  {author} {\bibinfo {author} {\bibfnamefont {W.-X.}\ \bibnamefont
  {Li}}, \bibinfo {author} {\bibfnamefont {C.}~\bibnamefont {Stampfl}},\ and\
  \bibinfo {author} {\bibfnamefont {M.}~\bibnamefont {Scheffler}},\ }\href
  {https://doi.org/10.1103/PhysRevB.65.075407} {\bibfield  {journal} {\bibinfo
  {journal} {Phys. Rev. B}\ }\textbf {\bibinfo {volume} {65}},\ \bibinfo
  {pages} {075407} (\bibinfo {year} {2002})}\BibitemShut {NoStop}%
\bibitem [{\citenamefont {Schintke}\ \emph {et~al.}(2001)\citenamefont
  {Schintke}, \citenamefont {Messerli}, \citenamefont {Morgenstern},
  \citenamefont {Nieminen},\ and\ \citenamefont {Schneider}}]{Schintke2001}%
  \BibitemOpen
  \bibfield  {author} {\bibinfo {author} {\bibfnamefont {S.}~\bibnamefont
  {Schintke}}, \bibinfo {author} {\bibfnamefont {S.}~\bibnamefont {Messerli}},
  \bibinfo {author} {\bibfnamefont {K.}~\bibnamefont {Morgenstern}}, \bibinfo
  {author} {\bibfnamefont {J.}~\bibnamefont {Nieminen}},\ and\ \bibinfo
  {author} {\bibfnamefont {W.-D.}\ \bibnamefont {Schneider}},\ }\href
  {https://doi.org/10.1063/1.1346687} {\bibfield  {journal} {\bibinfo
  {journal} {J. Chem. Phys.}\ }\textbf {\bibinfo {volume} {114}},\ \bibinfo
  {pages} {4206} (\bibinfo {year} {2001})}\BibitemShut {NoStop}%
\bibitem [{\citenamefont {S{\'{a}}nchez}\ \emph {et~al.}(2024)\citenamefont
  {S{\'{a}}nchez}, \citenamefont {Caporale}, \citenamefont {Degtev},
  \citenamefont {{Di Gaspare}}, \citenamefont {Persichetti}, \citenamefont
  {Sansotera}, \citenamefont {{G{\'{o}}mez Pueyo}}, \citenamefont {{De Seta}},
  \citenamefont {Lorenzana},\ and\ \citenamefont {Camilli}}]{Sanchez2024}%
  \BibitemOpen
  \bibfield  {author} {\bibinfo {author} {\bibfnamefont {J.~A.}\ \bibnamefont
  {S{\'{a}}nchez}}, \bibinfo {author} {\bibfnamefont {A.}~\bibnamefont
  {Caporale}}, \bibinfo {author} {\bibfnamefont {I.}~\bibnamefont {Degtev}},
  \bibinfo {author} {\bibfnamefont {L.}~\bibnamefont {{Di Gaspare}}}, \bibinfo
  {author} {\bibfnamefont {L.}~\bibnamefont {Persichetti}}, \bibinfo {author}
  {\bibfnamefont {M.}~\bibnamefont {Sansotera}}, \bibinfo {author}
  {\bibfnamefont {A.}~\bibnamefont {{G{\'{o}}mez Pueyo}}}, \bibinfo {author}
  {\bibfnamefont {M.}~\bibnamefont {{De Seta}}}, \bibinfo {author}
  {\bibfnamefont {J.}~\bibnamefont {Lorenzana}},\ and\ \bibinfo {author}
  {\bibfnamefont {L.}~\bibnamefont {Camilli}},\ }\href@noop {} {\bibinfo
  {title} {{The initial stages of silver fluorination, arXiv:2410.04858}}}
  (\bibinfo {year} {2024})\BibitemShut {NoStop}%
\bibitem [{\citenamefont {Lang}(1986)}]{Lang1986}%
  \BibitemOpen
  \bibfield  {author} {\bibinfo {author} {\bibfnamefont {N.~D.}\ \bibnamefont
  {Lang}},\ }\href {https://doi.org/10.1103/PhysRevB.34.5947} {\bibfield
  {journal} {\bibinfo  {journal} {Phys. Rev. B}\ }\textbf {\bibinfo {volume}
  {34}},\ \bibinfo {pages} {5947} (\bibinfo {year} {1986})}\BibitemShut
  {NoStop}%
\bibitem [{\citenamefont {Lang}(1987)}]{Lang1987}%
  \BibitemOpen
  \bibfield  {author} {\bibinfo {author} {\bibfnamefont {N.~D.}\ \bibnamefont
  {Lang}},\ }\href {https://doi.org/10.1103/PhysRevLett.58.45} {\bibfield
  {journal} {\bibinfo  {journal} {Phys. Rev. Lett.}\ }\textbf {\bibinfo
  {volume} {58}},\ \bibinfo {pages} {45} (\bibinfo {year} {1987})}\BibitemShut
  {NoStop}%
\bibitem [{\citenamefont {Sautet}(1997)}]{Sautet1997}%
  \BibitemOpen
  \bibfield  {author} {\bibinfo {author} {\bibfnamefont {P.}~\bibnamefont
  {Sautet}},\ }\href {https://doi.org/10.1016/S0039-6028(96)01239-3} {\bibfield
   {journal} {\bibinfo  {journal} {Surf. Sci.}\ }\textbf {\bibinfo {volume}
  {374}},\ \bibinfo {pages} {406} (\bibinfo {year} {1997})}\BibitemShut
  {NoStop}%
\bibitem [{\citenamefont {Tersoff}\ and\ \citenamefont
  {Hamann}(1983)}]{Tersoff1983}%
  \BibitemOpen
  \bibfield  {author} {\bibinfo {author} {\bibfnamefont {J.}~\bibnamefont
  {Tersoff}}\ and\ \bibinfo {author} {\bibfnamefont {D.~R.}\ \bibnamefont
  {Hamann}},\ }\href {https://doi.org/10.1103/PhysRevLett.50.1998} {\bibfield
  {journal} {\bibinfo  {journal} {Phys. Rev. Lett.}\ }\textbf {\bibinfo
  {volume} {50}},\ \bibinfo {pages} {1998} (\bibinfo {year}
  {1983})}\BibitemShut {NoStop}%
\bibitem [{\citenamefont {Otero-de-la Roza}\ \emph {et~al.}(2009)\citenamefont
  {Otero-de-la Roza}, \citenamefont {Blanco}, \citenamefont {Pend{\'{a}}s},\
  and\ \citenamefont {Lua{\~{n}}a}}]{OTERODELAROZA2009}%
  \BibitemOpen
  \bibfield  {author} {\bibinfo {author} {\bibfnamefont {A.}~\bibnamefont
  {Otero-de-la Roza}}, \bibinfo {author} {\bibfnamefont {M.~A.}\ \bibnamefont
  {Blanco}}, \bibinfo {author} {\bibfnamefont {A.~M.}\ \bibnamefont
  {Pend{\'{a}}s}},\ and\ \bibinfo {author} {\bibfnamefont {V.}~\bibnamefont
  {Lua{\~{n}}a}},\ }\href
  {https://doi.org/https://doi.org/10.1016/j.cpc.2008.07.018} {\bibfield
  {journal} {\bibinfo  {journal} {Comput. Phys. Commun.}\ }\textbf {\bibinfo
  {volume} {180}},\ \bibinfo {pages} {157} (\bibinfo {year}
  {2009})}\BibitemShut {NoStop}%
\bibitem [{\citenamefont {Otero-de-la Roza}\ \emph {et~al.}(2014)\citenamefont
  {Otero-de-la Roza}, \citenamefont {Johnson},\ and\ \citenamefont
  {Lua{\~{n}}a}}]{OTERODELAROZA20141007}%
  \BibitemOpen
  \bibfield  {author} {\bibinfo {author} {\bibfnamefont {A.}~\bibnamefont
  {Otero-de-la Roza}}, \bibinfo {author} {\bibfnamefont {E.~R.}\ \bibnamefont
  {Johnson}},\ and\ \bibinfo {author} {\bibfnamefont {V.}~\bibnamefont
  {Lua{\~{n}}a}},\ }\href
  {https://doi.org/https://doi.org/10.1016/j.cpc.2013.10.026} {\bibfield
  {journal} {\bibinfo  {journal} {Comput. Phys. Commun.}\ }\textbf {\bibinfo
  {volume} {185}},\ \bibinfo {pages} {1007} (\bibinfo {year}
  {2014})}\BibitemShut {NoStop}%
\bibitem [{\citenamefont {Li}\ \emph {et~al.}(2003)\citenamefont {Li},
  \citenamefont {Stampfl},\ and\ \citenamefont {Scheffler}}]{Li2003a}%
  \BibitemOpen
  \bibfield  {author} {\bibinfo {author} {\bibfnamefont {W.-X.~X.}\
  \bibnamefont {Li}}, \bibinfo {author} {\bibfnamefont {C.}~\bibnamefont
  {Stampfl}},\ and\ \bibinfo {author} {\bibfnamefont {M.}~\bibnamefont
  {Scheffler}},\ }\href {https://doi.org/10.1103/PhysRevB.68.165412} {\bibfield
   {journal} {\bibinfo  {journal} {Phys. Rev. B}\ }\textbf {\bibinfo {volume}
  {68}},\ \bibinfo {pages} {165412} (\bibinfo {year} {2003})},\ \Eprint
  {https://arxiv.org/abs/0305312} {arXiv:0305312 [cond-mat]} \BibitemShut
  {NoStop}%
\bibitem [{\citenamefont {Reuter}\ and\ \citenamefont
  {Scheffler}(2001)}]{Reuter2001}%
  \BibitemOpen
  \bibfield  {author} {\bibinfo {author} {\bibfnamefont {K.}~\bibnamefont
  {Reuter}}\ and\ \bibinfo {author} {\bibfnamefont {M.}~\bibnamefont
  {Scheffler}},\ }\href {https://doi.org/10.1103/PhysRevB.65.035406} {\bibfield
   {journal} {\bibinfo  {journal} {Phys. Rev. B}\ }\textbf {\bibinfo {volume}
  {65}},\ \bibinfo {pages} {035406} (\bibinfo {year} {2001})}\BibitemShut
  {NoStop}%
\bibitem [{\citenamefont {Kresse}\ and\ \citenamefont {Hafner}(1993)}]{VASP1}%
  \BibitemOpen
  \bibfield  {author} {\bibinfo {author} {\bibfnamefont {G.}~\bibnamefont
  {Kresse}}\ and\ \bibinfo {author} {\bibfnamefont {J.}~\bibnamefont
  {Hafner}},\ }\href {https://doi.org/10.1103/PhysRevB.47.558} {\bibfield
  {journal} {\bibinfo  {journal} {Phys. Rev. B}\ }\textbf {\bibinfo {volume}
  {47}},\ \bibinfo {pages} {558} (\bibinfo {year} {1993})}\BibitemShut
  {NoStop}%
\bibitem [{\citenamefont {Kresse}\ and\ \citenamefont
  {Furthm{\"{u}}ller}(1996{\natexlab{a}})}]{VASP2}%
  \BibitemOpen
  \bibfield  {author} {\bibinfo {author} {\bibfnamefont {G.}~\bibnamefont
  {Kresse}}\ and\ \bibinfo {author} {\bibfnamefont {J.}~\bibnamefont
  {Furthm{\"{u}}ller}},\ }\href
  {https://doi.org/https://doi.org/10.1016/0927-0256(96)00008-0} {\bibfield
  {journal} {\bibinfo  {journal} {Comput. Mater. Sci.}\ }\textbf {\bibinfo
  {volume} {6}},\ \bibinfo {pages} {15} (\bibinfo {year}
  {1996}{\natexlab{a}})}\BibitemShut {NoStop}%
\bibitem [{\citenamefont {Kresse}\ and\ \citenamefont
  {Furthm{\"{u}}ller}(1996{\natexlab{b}})}]{VASP3}%
  \BibitemOpen
  \bibfield  {author} {\bibinfo {author} {\bibfnamefont {G.}~\bibnamefont
  {Kresse}}\ and\ \bibinfo {author} {\bibfnamefont {J.}~\bibnamefont
  {Furthm{\"{u}}ller}},\ }\href {https://doi.org/10.1103/PhysRevB.54.11169}
  {\bibfield  {journal} {\bibinfo  {journal} {Phys. Rev. B}\ }\textbf {\bibinfo
  {volume} {54}},\ \bibinfo {pages} {11169} (\bibinfo {year}
  {1996}{\natexlab{b}})}\BibitemShut {NoStop}%
\bibitem [{\citenamefont {Kresse}\ and\ \citenamefont
  {Joubert}(1999)}]{VASPPAW}%
  \BibitemOpen
  \bibfield  {author} {\bibinfo {author} {\bibfnamefont {G.}~\bibnamefont
  {Kresse}}\ and\ \bibinfo {author} {\bibfnamefont {D.}~\bibnamefont
  {Joubert}},\ }\href {https://doi.org/10.1103/PhysRevB.59.1758} {\bibfield
  {journal} {\bibinfo  {journal} {Phys. Rev. B}\ }\textbf {\bibinfo {volume}
  {59}},\ \bibinfo {pages} {1758} (\bibinfo {year} {1999})}\BibitemShut
  {NoStop}%
\bibitem [{\citenamefont {Perdew}\ \emph {et~al.}(2008)\citenamefont {Perdew},
  \citenamefont {Ruzsinszky}, \citenamefont {Csonka}, \citenamefont {Vydrov},
  \citenamefont {Scuseria}, \citenamefont {Constantin}, \citenamefont {Zhou},\
  and\ \citenamefont {Burke}}]{Perdew2008}%
  \BibitemOpen
  \bibfield  {author} {\bibinfo {author} {\bibfnamefont {J.~P.}\ \bibnamefont
  {Perdew}}, \bibinfo {author} {\bibfnamefont {A.}~\bibnamefont {Ruzsinszky}},
  \bibinfo {author} {\bibfnamefont {G.~I.}\ \bibnamefont {Csonka}}, \bibinfo
  {author} {\bibfnamefont {O.~A.}\ \bibnamefont {Vydrov}}, \bibinfo {author}
  {\bibfnamefont {G.~E.}\ \bibnamefont {Scuseria}}, \bibinfo {author}
  {\bibfnamefont {L.~A.}\ \bibnamefont {Constantin}}, \bibinfo {author}
  {\bibfnamefont {X.}~\bibnamefont {Zhou}},\ and\ \bibinfo {author}
  {\bibfnamefont {K.}~\bibnamefont {Burke}},\ }\href
  {https://doi.org/10.1103/PhysRevLett.100.136406} {\bibfield  {journal}
  {\bibinfo  {journal} {Phys. Rev. Lett.}\ }\textbf {\bibinfo {volume} {100}},\
  \bibinfo {pages} {136406} (\bibinfo {year} {2008})}\BibitemShut {NoStop}%
\bibitem [{\citenamefont {King}(1981)}]{King1981}%
  \BibitemOpen
  \bibfield  {author} {\bibinfo {author} {\bibfnamefont {H.~W.}\ \bibnamefont
  {King}},\ }\href {https://doi.org/10.1007/BF02868307} {\bibfield  {journal}
  {\bibinfo  {journal} {Bull. Alloy Phase Diagrams}\ }\textbf {\bibinfo
  {volume} {2}},\ \bibinfo {pages} {401} (\bibinfo {year} {1981})}\BibitemShut
  {NoStop}%
\bibitem [{\citenamefont {Hu}\ \emph {et~al.}(2005)\citenamefont {Hu},
  \citenamefont {Cai}, \citenamefont {Li}, \citenamefont {Gan},\ and\
  \citenamefont {Chen}}]{Hu2005}%
  \BibitemOpen
  \bibfield  {author} {\bibinfo {author} {\bibfnamefont {J.}~\bibnamefont
  {Hu}}, \bibinfo {author} {\bibfnamefont {W.}~\bibnamefont {Cai}}, \bibinfo
  {author} {\bibfnamefont {C.}~\bibnamefont {Li}}, \bibinfo {author}
  {\bibfnamefont {Y.}~\bibnamefont {Gan}},\ and\ \bibinfo {author}
  {\bibfnamefont {L.}~\bibnamefont {Chen}},\ }\href
  {https://doi.org/10.1063/1.1901803} {\bibfield  {journal} {\bibinfo
  {journal} {Applied Physics Letters}\ }\textbf {\bibinfo {volume} {86}},\
  \bibinfo {pages} {151915} (\bibinfo {year} {2005})},\ \Eprint
  {https://arxiv.org/abs/https://pubs.aip.org/aip/apl/article-pdf/doi/10.1063/1.1901803/14332160/151915\_1\_online.pdf}
  {https://pubs.aip.org/aip/apl/article-pdf/doi/10.1063/1.1901803/14332160/151915\_1\_online.pdf}
  \BibitemShut {NoStop}%
\bibitem [{\citenamefont {Patra}\ \emph {et~al.}(2017)\citenamefont {Patra},
  \citenamefont {Bates}, \citenamefont {Sun},\ and\ \citenamefont
  {Perdew}}]{Perdew20172}%
  \BibitemOpen
  \bibfield  {author} {\bibinfo {author} {\bibfnamefont {A.}~\bibnamefont
  {Patra}}, \bibinfo {author} {\bibfnamefont {J.~E.}\ \bibnamefont {Bates}},
  \bibinfo {author} {\bibfnamefont {J.}~\bibnamefont {Sun}},\ and\ \bibinfo
  {author} {\bibfnamefont {J.~P.}\ \bibnamefont {Perdew}},\ }\href
  {https://doi.org/10.1073/pnas.1713320114} {\bibfield  {journal} {\bibinfo
  {journal} {Proc. Natl. Acad. Sci.}\ }\textbf {\bibinfo {volume} {114}},\
  \bibinfo {pages} {E9188} (\bibinfo {year} {2017})}\BibitemShut {NoStop}%
\bibitem [{\citenamefont {Banerjee}\ \emph {et~al.}(2018)\citenamefont
  {Banerjee}, \citenamefont {Behnle}, \citenamefont {Galbraith}, \citenamefont
  {Settels}, \citenamefont {Engels}, \citenamefont {Tonner},\ and\
  \citenamefont {Fink}}]{Banerjee2018}%
  \BibitemOpen
  \bibfield  {author} {\bibinfo {author} {\bibfnamefont {J.}~\bibnamefont
  {Banerjee}}, \bibinfo {author} {\bibfnamefont {S.}~\bibnamefont {Behnle}},
  \bibinfo {author} {\bibfnamefont {M.~C.~E.}\ \bibnamefont {Galbraith}},
  \bibinfo {author} {\bibfnamefont {V.}~\bibnamefont {Settels}}, \bibinfo
  {author} {\bibfnamefont {B.}~\bibnamefont {Engels}}, \bibinfo {author}
  {\bibfnamefont {R.}~\bibnamefont {Tonner}},\ and\ \bibinfo {author}
  {\bibfnamefont {R.~F.}\ \bibnamefont {Fink}},\ }\href
  {https://doi.org/https://doi.org/10.1002/jcc.25159} {\bibfield  {journal}
  {\bibinfo  {journal} {J. Comput. Chem.}\ }\textbf {\bibinfo {volume} {39}},\
  \bibinfo {pages} {844} (\bibinfo {year} {2018})}\BibitemShut {NoStop}%
\bibitem [{\citenamefont {Neugebauer}\ and\ \citenamefont
  {Scheffler}(1992)}]{Neugebauer1992}%
  \BibitemOpen
  \bibfield  {author} {\bibinfo {author} {\bibfnamefont {J.}~\bibnamefont
  {Neugebauer}}\ and\ \bibinfo {author} {\bibfnamefont {M.}~\bibnamefont
  {Scheffler}},\ }\href {https://doi.org/10.1103/PhysRevB.46.16067} {\bibfield
  {journal} {\bibinfo  {journal} {Phys. Rev. B}\ }\textbf {\bibinfo {volume}
  {46}},\ \bibinfo {pages} {16067} (\bibinfo {year} {1992})}\BibitemShut
  {NoStop}%
\bibitem [{\citenamefont {Makov}\ and\ \citenamefont
  {Payne}(1995)}]{Makov1995}%
  \BibitemOpen
  \bibfield  {author} {\bibinfo {author} {\bibfnamefont {G.}~\bibnamefont
  {Makov}}\ and\ \bibinfo {author} {\bibfnamefont {M.~C.}\ \bibnamefont
  {Payne}},\ }\href {https://doi.org/10.1103/PhysRevB.51.4014} {\bibfield
  {journal} {\bibinfo  {journal} {Phys. Rev. B}\ }\textbf {\bibinfo {volume}
  {51}},\ \bibinfo {pages} {4014} (\bibinfo {year} {1995})}\BibitemShut
  {NoStop}%
\bibitem [{\citenamefont {Migani}\ and\ \citenamefont
  {Illas}(2006)}]{Migani2006}%
  \BibitemOpen
  \bibfield  {author} {\bibinfo {author} {\bibfnamefont {A.}~\bibnamefont
  {Migani}}\ and\ \bibinfo {author} {\bibfnamefont {F.}~\bibnamefont {Illas}},\
  }\href {https://doi.org/10.1021/jp060400u} {\bibfield  {journal} {\bibinfo
  {journal} {J. Phys. Chem. B}\ }\textbf {\bibinfo {volume} {110}},\ \bibinfo
  {pages} {11894} (\bibinfo {year} {2006})}\BibitemShut {NoStop}%
\bibitem [{\citenamefont {Bl\"ochl}(1994)}]{Blochl1994}%
  \BibitemOpen
  \bibfield  {author} {\bibinfo {author} {\bibfnamefont {P.~E.}\ \bibnamefont
  {Bl\"ochl}},\ }\href {https://doi.org/10.1103/PhysRevB.50.17953} {\bibfield
  {journal} {\bibinfo  {journal} {Phys. Rev. B}\ }\textbf {\bibinfo {volume}
  {50}},\ \bibinfo {pages} {17953} (\bibinfo {year} {1994})}\BibitemShut
  {NoStop}%
\bibitem [{\citenamefont {Dell'Anna}\ \emph {et~al.}(2005)\citenamefont
  {Dell'Anna}, \citenamefont {Lorenzana}, \citenamefont {Capone}, \citenamefont
  {Castellani},\ and\ \citenamefont {Grilli}}]{DellAnna2005}%
  \BibitemOpen
  \bibfield  {author} {\bibinfo {author} {\bibfnamefont {L.}~\bibnamefont
  {Dell'Anna}}, \bibinfo {author} {\bibfnamefont {J.}~\bibnamefont
  {Lorenzana}}, \bibinfo {author} {\bibfnamefont {M.}~\bibnamefont {Capone}},
  \bibinfo {author} {\bibfnamefont {C.}~\bibnamefont {Castellani}},\ and\
  \bibinfo {author} {\bibfnamefont {M.}~\bibnamefont {Grilli}},\ }\href
  {https://doi.org/10.1103/PhysRevB.71.064518} {\bibfield  {journal} {\bibinfo
  {journal} {Phys. Rev. B - Condens. Matter Mater. Phys.}\ }\textbf {\bibinfo
  {volume} {71}},\ \bibinfo {pages} {064518} (\bibinfo {year} {2005})},\
  \Eprint {https://arxiv.org/abs/0407028} {arXiv:0407028 [cond-mat]}
  \BibitemShut {NoStop}%
\bibitem [{\citenamefont {Kreisel}\ \emph {et~al.}(2016)\citenamefont
  {Kreisel}, \citenamefont {Nelson}, \citenamefont {Berlijn}, \citenamefont
  {Ku}, \citenamefont {Aluru}, \citenamefont {Chi}, \citenamefont {Zhou},
  \citenamefont {Singh}, \citenamefont {Wahl}, \citenamefont {Liang},
  \citenamefont {Hardy}, \citenamefont {Bonn}, \citenamefont {Hirschfeld},\
  and\ \citenamefont {Andersen}}]{Kreisel2016}%
  \BibitemOpen
  \bibfield  {author} {\bibinfo {author} {\bibfnamefont {A.}~\bibnamefont
  {Kreisel}}, \bibinfo {author} {\bibfnamefont {R.}~\bibnamefont {Nelson}},
  \bibinfo {author} {\bibfnamefont {T.}~\bibnamefont {Berlijn}}, \bibinfo
  {author} {\bibfnamefont {W.}~\bibnamefont {Ku}}, \bibinfo {author}
  {\bibfnamefont {R.}~\bibnamefont {Aluru}}, \bibinfo {author} {\bibfnamefont
  {S.}~\bibnamefont {Chi}}, \bibinfo {author} {\bibfnamefont {H.}~\bibnamefont
  {Zhou}}, \bibinfo {author} {\bibfnamefont {U.~R.}\ \bibnamefont {Singh}},
  \bibinfo {author} {\bibfnamefont {P.}~\bibnamefont {Wahl}}, \bibinfo {author}
  {\bibfnamefont {R.}~\bibnamefont {Liang}}, \bibinfo {author} {\bibfnamefont
  {W.~N.}\ \bibnamefont {Hardy}}, \bibinfo {author} {\bibfnamefont {D.~A.}\
  \bibnamefont {Bonn}}, \bibinfo {author} {\bibfnamefont {P.~J.}\ \bibnamefont
  {Hirschfeld}},\ and\ \bibinfo {author} {\bibfnamefont {B.~M.}\ \bibnamefont
  {Andersen}},\ }\href {https://doi.org/10.1103/PhysRevB.94.224518} {\bibfield
  {journal} {\bibinfo  {journal} {Phys. Rev. B}\ }\textbf {\bibinfo {volume}
  {94}},\ \bibinfo {pages} {224518} (\bibinfo {year} {2016})}\BibitemShut
  {NoStop}%
\bibitem [{\citenamefont {Perdew}\ \emph {et~al.}(1996)\citenamefont {Perdew},
  \citenamefont {Burke},\ and\ \citenamefont {Ernzerhof}}]{Perdew1996}%
  \BibitemOpen
  \bibfield  {author} {\bibinfo {author} {\bibfnamefont {J.~P.}\ \bibnamefont
  {Perdew}}, \bibinfo {author} {\bibfnamefont {K.}~\bibnamefont {Burke}},\ and\
  \bibinfo {author} {\bibfnamefont {M.}~\bibnamefont {Ernzerhof}},\ }\href
  {https://doi.org/10.1103/PhysRevLett.77.3865} {\bibfield  {journal} {\bibinfo
   {journal} {Phys. Rev. Lett.}\ }\textbf {\bibinfo {volume} {77}},\ \bibinfo
  {pages} {3865} (\bibinfo {year} {1996})}\BibitemShut {NoStop}%
\bibitem [{\citenamefont {Matth{\'\i}asson}\ \emph {et~al.}(2021)\citenamefont
  {Matth{\'\i}asson}, \citenamefont {Kvaran}, \citenamefont {Garcia},
  \citenamefont {Weidner},\ and\ \citenamefont
  {Szt{\'a}ray}}]{Matthiasson2021}%
  \BibitemOpen
  \bibfield  {author} {\bibinfo {author} {\bibfnamefont {K.}~\bibnamefont
  {Matth{\'\i}asson}}, \bibinfo {author} {\bibfnamefont {{\'A}.}~\bibnamefont
  {Kvaran}}, \bibinfo {author} {\bibfnamefont {G.~A.}\ \bibnamefont {Garcia}},
  \bibinfo {author} {\bibfnamefont {P.}~\bibnamefont {Weidner}},\ and\ \bibinfo
  {author} {\bibfnamefont {B.}~\bibnamefont {Szt{\'a}ray}},\ }\href
  {https://doi.org/10.1039/D1CP00140J} {\bibfield  {journal} {\bibinfo
  {journal} {Physical Chemistry Chemical Physics}\ }\textbf {\bibinfo {volume}
  {23}},\ \bibinfo {pages} {8292} (\bibinfo {year} {2021})}\BibitemShut
  {NoStop}%
\bibitem [{\citenamefont {DeCorpo}\ \emph {et~al.}(1970)\citenamefont
  {DeCorpo}, \citenamefont {Steiger}, \citenamefont {Franklin},\ and\
  \citenamefont {Margrave}}]{Decorpo1970}%
  \BibitemOpen
  \bibfield  {author} {\bibinfo {author} {\bibfnamefont {J.~J.}\ \bibnamefont
  {DeCorpo}}, \bibinfo {author} {\bibfnamefont {R.~P.}\ \bibnamefont
  {Steiger}}, \bibinfo {author} {\bibfnamefont {J.~L.}\ \bibnamefont
  {Franklin}},\ and\ \bibinfo {author} {\bibfnamefont {J.~L.}\ \bibnamefont
  {Margrave}},\ }\href {https://doi.org/10.1063/1.1674160} {\bibfield
  {journal} {\bibinfo  {journal} {J. Chem. Phys.}\ }\textbf {\bibinfo {volume}
  {53}},\ \bibinfo {pages} {936} (\bibinfo {year} {1970})}\BibitemShut
  {NoStop}%
\bibitem [{Note1()}]{Note1}%
  \BibitemOpen
  \bibinfo {note} {Since our computation does not take into account the
  zero-point vibrational energy, it is more fair to remove the vibrational
  contribution~\cite {Pople1989} from the experiment, which yields $E^{\protect
  \rm exp}_d=0.855$ eV, still significantly smaller than the
  theory}\BibitemShut {NoStop}%
\bibitem [{\citenamefont {Wang}\ \emph {et~al.}(2001)\citenamefont {Wang},
  \citenamefont {Xu}, \citenamefont {Kakeshita}, \citenamefont {Uchida},
  \citenamefont {Ono}, \citenamefont {Ando},\ and\ \citenamefont
  {Ong}}]{Wang2001}%
  \BibitemOpen
  \bibfield  {author} {\bibinfo {author} {\bibfnamefont {Y.}~\bibnamefont
  {Wang}}, \bibinfo {author} {\bibfnamefont {Z.~a.}\ \bibnamefont {Xu}},
  \bibinfo {author} {\bibfnamefont {T.}~\bibnamefont {Kakeshita}}, \bibinfo
  {author} {\bibfnamefont {S.}~\bibnamefont {Uchida}}, \bibinfo {author}
  {\bibfnamefont {S.}~\bibnamefont {Ono}}, \bibinfo {author} {\bibfnamefont
  {Y.}~\bibnamefont {Ando}},\ and\ \bibinfo {author} {\bibfnamefont {N.~P.}\
  \bibnamefont {Ong}},\ }\href {https://doi.org/10.1103/PhysRevB.64.224519}
  {\bibfield  {journal} {\bibinfo  {journal} {Phys. Rev. B}\ }\textbf {\bibinfo
  {volume} {64}},\ \bibinfo {pages} {224519} (\bibinfo {year}
  {2001})}\BibitemShut {NoStop}%
\bibitem [{\citenamefont {Scheffler}\ and\ \citenamefont
  {Stampfl}(2000)}]{Scheffler2000}%
  \BibitemOpen
  \bibfield  {author} {\bibinfo {author} {\bibfnamefont {M.}~\bibnamefont
  {Scheffler}}\ and\ \bibinfo {author} {\bibfnamefont {C.}~\bibnamefont
  {Stampfl}},\ }in\ \href@noop {} {{\bibinfo {booktitle} {\em Handb. Surf.
  Sciene - Electron. Struct.}}},\ \bibinfo {editor} {edited by\ \bibinfo
  {editor} {\bibfnamefont {K.}~\bibnamefont {Horn}}\ and\ \bibinfo {editor}
  {\bibfnamefont {M.}~\bibnamefont {Scheffler}}}\ (\bibinfo  {publisher}
  {North-Holland},\ \bibinfo {address} {Amsterdam},\ \bibinfo {year} {2000})\
  p.\ \bibinfo {pages} {285}\BibitemShut {NoStop}%
\bibitem [{\citenamefont {Linstrom}\ and\ \citenamefont
  {Mallard}(2023)}]{Linstrom2023}%
  \BibitemOpen
  \bibinfo {editor} {\bibfnamefont {P.}~\bibnamefont {Linstrom}}\ and\ \bibinfo
  {editor} {\bibfnamefont {W.}~\bibnamefont {Mallard}},\ eds.,\ \href
  {https://doi.org/10.18434/T4D303} { {\bibinfo {title} {\em NIST Chemistry
  WebBook, NIST Standard Reference Database Number 69}}}\ (\bibinfo
  {publisher} {National Institute of Standards and Technology},\ \bibinfo
  {address} {Gaithersburg MD},\ \bibinfo {year} {2023})\BibitemShut {NoStop}%
\bibitem [{\citenamefont {Cabrera}\ and\ \citenamefont
  {Mott}(1949)}]{Cabrera1949}%
  \BibitemOpen
  \bibfield  {author} {\bibinfo {author} {\bibfnamefont {N.}~\bibnamefont
  {Cabrera}}\ and\ \bibinfo {author} {\bibfnamefont {N.~F.}\ \bibnamefont
  {Mott}},\ }\href {https://doi.org/10.1088/0034-4885/12/1/308} {\bibfield
  {journal} {\bibinfo  {journal} {Reports Prog. Phys.}\ }\textbf {\bibinfo
  {volume} {12}},\ \bibinfo {pages} {308} (\bibinfo {year} {1949})}\BibitemShut
  {NoStop}%
\bibitem [{\citenamefont {Qiu}\ and\ \citenamefont {Yarmoff}(2001)}]{Qiu2001}%
  \BibitemOpen
  \bibfield  {author} {\bibinfo {author} {\bibfnamefont {S.~R.}\ \bibnamefont
  {Qiu}}\ and\ \bibinfo {author} {\bibfnamefont {J.~A.}\ \bibnamefont
  {Yarmoff}},\ }\href {https://doi.org/10.1103/PhysRevB.63.115409} {\bibfield
  {journal} {\bibinfo  {journal} {Phys. Rev. B}\ }\textbf {\bibinfo {volume}
  {63}},\ \bibinfo {pages} {115409} (\bibinfo {year} {2001})}\BibitemShut
  {NoStop}%
\bibitem [{Note2()}]{Note2}%
  \BibitemOpen
  \bibinfo {note} {See Supplemental Material at [URL will be inserted by
  publisher] for all the ATs computed and the corresponding reference values of
  the apparent height of the tip.}\BibitemShut {Stop}%
\bibitem [{\citenamefont {Kohn}\ and\ \citenamefont {Sham}(1965)}]{Kohn1965}%
  \BibitemOpen
  \bibfield  {author} {\bibinfo {author} {\bibfnamefont {W.}~\bibnamefont
  {Kohn}}\ and\ \bibinfo {author} {\bibfnamefont {L.~J.}\ \bibnamefont
  {Sham}},\ }\href {https://doi.org/10.1103/PhysRev.140.A1133} {\bibfield
  {journal} {\bibinfo  {journal} {Phys. Rev.}\ }\textbf {\bibinfo {volume}
  {140}},\ \bibinfo {pages} {A1133} (\bibinfo {year} {1965})}\BibitemShut
  {NoStop}%
\bibitem [{\citenamefont {Schott}\ and\ \citenamefont
  {White}(1994)}]{Schott1994}%
  \BibitemOpen
  \bibfield  {author} {\bibinfo {author} {\bibfnamefont {J.~H.}\ \bibnamefont
  {Schott}}\ and\ \bibinfo {author} {\bibfnamefont {H.~S.}\ \bibnamefont
  {White}},\ }\href {https://doi.org/10.1021/j100052a050} {\bibfield  {journal}
  {\bibinfo  {journal} {J. Phys. Chem.}\ }\textbf {\bibinfo {volume} {98}},\
  \bibinfo {pages} {297} (\bibinfo {year} {1994})}\BibitemShut {NoStop}%
\bibitem [{\citenamefont {Slater}(1930)}]{Slater1930}%
  \BibitemOpen
  \bibfield  {author} {\bibinfo {author} {\bibfnamefont {J.~C.}\ \bibnamefont
  {Slater}},\ }\href {https://doi.org/10.1103/PhysRev.36.57} {\bibfield
  {journal} {\bibinfo  {journal} {Phys. Rev.}\ }\textbf {\bibinfo {volume}
  {36}},\ \bibinfo {pages} {57} (\bibinfo {year} {1930})}\BibitemShut {NoStop}%
\bibitem [{\citenamefont {Clementi}\ and\ \citenamefont
  {Raimondi}(1963)}]{Clementi1963}%
  \BibitemOpen
  \bibfield  {author} {\bibinfo {author} {\bibfnamefont {E.}~\bibnamefont
  {Clementi}}\ and\ \bibinfo {author} {\bibfnamefont {D.~L.}\ \bibnamefont
  {Raimondi}},\ }\href {https://doi.org/10.1063/1.1733573} {\bibfield
  {journal} {\bibinfo  {journal} {J. Chem. Phys.}\ }\textbf {\bibinfo {volume}
  {38}},\ \bibinfo {pages} {2686} (\bibinfo {year} {1963})}\BibitemShut
  {NoStop}%
\bibitem [{\citenamefont {Mann}(1968)}]{Mann1968}%
  \BibitemOpen
  \bibfield  {author} {\bibinfo {author} {\bibfnamefont {J.~B.}\ \bibnamefont
  {Mann}},\ }\href@noop {} { {\bibinfo {title} {{\em Atomic Structure
  Calculations II. Hartree-Fock Wave functions and Radial Expectation Values:
  Hydrogen to Lawrencium, Report LA-3691}}}},\ \bibinfo {type} {Tech. Rep.}\
  (\bibinfo  {institution} {Los Alamos Scientific Laboratory},\ \bibinfo
  {address} {Los Alamos, New Mexico},\ \bibinfo {year} {1968})\BibitemShut
  {NoStop}%
\bibitem [{\citenamefont {Slater}(1964)}]{Slater1964}%
  \BibitemOpen
  \bibfield  {author} {\bibinfo {author} {\bibfnamefont {J.~C.}\ \bibnamefont
  {Slater}},\ }\href {https://doi.org/10.1063/1.1725697} {\bibfield  {journal}
  {\bibinfo  {journal} {J. Chem. Phys.}\ }\textbf {\bibinfo {volume} {41}},\
  \bibinfo {pages} {3199} (\bibinfo {year} {1964})}\BibitemShut {NoStop}%
\bibitem [{\citenamefont {Shannon}(1976)}]{Shannon1976}%
  \BibitemOpen
  \bibfield  {author} {\bibinfo {author} {\bibfnamefont {R.~D.}\ \bibnamefont
  {Shannon}},\ }\href {https://doi.org/10.1107/S0567739476001551} {\bibfield
  {journal} {\bibinfo  {journal} {Acta Crystallogr. Sect. A}\ }\textbf
  {\bibinfo {volume} {32}},\ \bibinfo {pages} {751} (\bibinfo {year}
  {1976})}\BibitemShut {NoStop}%
\bibitem [{\citenamefont {Sautet}\ and\ \citenamefont
  {Bocquet}(1996)}]{Sautet1996}%
  \BibitemOpen
  \bibfield  {author} {\bibinfo {author} {\bibfnamefont {P.}~\bibnamefont
  {Sautet}}\ and\ \bibinfo {author} {\bibfnamefont {M.}~\bibnamefont
  {Bocquet}},\ }\href {https://doi.org/10.1103/PhysRevB.53.4910} {\bibfield
  {journal} {\bibinfo  {journal} {Phys. Rev. B - Condens. Matter Mater. Phys.}\
  }\textbf {\bibinfo {volume} {53}},\ \bibinfo {pages} {4910} (\bibinfo {year}
  {1996})}\BibitemShut {NoStop}%
\bibitem [{\citenamefont {Filippetti}\ and\ \citenamefont
  {Fiorentini}(2009)}]{Filippetti2009}%
  \BibitemOpen
  \bibfield  {author} {\bibinfo {author} {\bibfnamefont {A.}~\bibnamefont
  {Filippetti}}\ and\ \bibinfo {author} {\bibfnamefont {V.}~\bibnamefont
  {Fiorentini}},\ }\href {https://doi.org/10.1140/epjb/e2009-00313-2}
  {\bibfield  {journal} {\bibinfo  {journal} {Eur. Phys. J. B}\ }\textbf
  {\bibinfo {volume} {71}},\ \bibinfo {pages} {139} (\bibinfo {year}
  {2009})}\BibitemShut {NoStop}%
\bibitem [{\citenamefont {Wintterlin}(2000)}]{Wintterlin2000}%
  \BibitemOpen
  \bibfield  {author} {\bibinfo {author} {\bibfnamefont {J.}~\bibnamefont
  {Wintterlin}},\ }\href {https://doi.org/10.1016/s0360-0564(02)45014-6}
  {\bibfield  {journal} {\bibinfo  {journal} {Adv. Catal.}\ }\textbf {\bibinfo
  {volume} {45}},\ \bibinfo {pages} {131} (\bibinfo {year} {2000})}\BibitemShut
  {NoStop}%
\bibitem [{\citenamefont {Jones}(1988)}]{Jones1988}%
  \BibitemOpen
  \bibfield  {author} {\bibinfo {author} {\bibfnamefont {R.~G.}\ \bibnamefont
  {Jones}},\ }\href
  {https://doi.org/https://doi.org/10.1016/0079-6816(88)90013-5} {\bibfield
  {journal} {\bibinfo  {journal} {Prog. Surf. Sci.}\ }\textbf {\bibinfo
  {volume} {27}},\ \bibinfo {pages} {25} (\bibinfo {year} {1988})}\BibitemShut
  {NoStop}%
\bibitem [{\citenamefont {Altman}(2001)}]{Altman2001}%
  \BibitemOpen
  \bibfield  {author} {\bibinfo {author} {\bibfnamefont {E.~I.}\ \bibnamefont
  {Altman}},\ }\href {https://doi.org/10.1007/10689660_27} {\bibinfo {title}
  {Halogens on metals and semiconductors: Datasheet from
  \uppercase{L}andolt-\uppercase{B}{\"o}rnstein - \uppercase{G}roup
  \uppercase{III} \uppercase{C}ondensed \uppercase{M}atter
  {\textperiodcentered} volume 42\uppercase{A} 1: ``\uppercase{A}dsorbed
  \uppercase{L}ayers on \uppercase{S}urfaces. \uppercase{P}art 1: Adsorption on
  \uppercase{S}urfaces and \uppercase{S}urface \uppercase{D}iffusion of
  \uppercase{A}dsorbates'' in \uppercase{S}pringer\uppercase{M}aterials
  (https://doi.org/10.1007/10689660{\_}27)}} (\bibinfo {year} {2001}),\
  \bibinfo {note} {\uppercase{C}opyright 2001 Springer-Verlag Berlin
  Heidelberg, Part of SpringerMaterials}\BibitemShut {NoStop}%
\bibitem [{\citenamefont {Serafin}\ \emph {et~al.}(1998)\citenamefont
  {Serafin}, \citenamefont {Liu},\ and\ \citenamefont
  {Seyedmonir}}]{Serafin1998}%
  \BibitemOpen
  \bibfield  {author} {\bibinfo {author} {\bibfnamefont {J.~G.}\ \bibnamefont
  {Serafin}}, \bibinfo {author} {\bibfnamefont {A.~C.}\ \bibnamefont {Liu}},\
  and\ \bibinfo {author} {\bibfnamefont {S.~R.}\ \bibnamefont {Seyedmonir}},\
  }\href {https://doi.org/https://doi.org/10.1016/S1381-1169(97)00263-X}
  {\bibfield  {journal} {\bibinfo  {journal} {J. Mol. Catal. A Chem.}\ }\textbf
  {\bibinfo {volume} {131}},\ \bibinfo {pages} {157} (\bibinfo {year}
  {1998})}\BibitemShut {NoStop}%
\bibitem [{\citenamefont {Pople}\ \emph {et~al.}(1989)\citenamefont {Pople},
  \citenamefont {Head‐Gordon}, \citenamefont {Fox}, \citenamefont
  {Raghavachari},\ and\ \citenamefont {Curtiss}}]{Pople1989}%
  \BibitemOpen
  \bibfield  {author} {\bibinfo {author} {\bibfnamefont {J.~A.}\ \bibnamefont
  {Pople}}, \bibinfo {author} {\bibfnamefont {M.}~\bibnamefont
  {Head‐Gordon}}, \bibinfo {author} {\bibfnamefont {D.~J.}\ \bibnamefont
  {Fox}}, \bibinfo {author} {\bibfnamefont {K.}~\bibnamefont {Raghavachari}},\
  and\ \bibinfo {author} {\bibfnamefont {L.~A.}\ \bibnamefont {Curtiss}},\
  }\href {https://doi.org/10.1063/1.456415} {\bibfield  {journal} {\bibinfo
  {journal} {J. Chem. Phys.}\ }\textbf {\bibinfo {volume} {90}},\ \bibinfo
  {pages} {5622} (\bibinfo {year} {1989})}\BibitemShut {NoStop}%
\end{thebibliography}
\providecommand{\noopsort}[1]{}\providecommand{\singleletter}[1]{#1}%

\end{document}